\RequirePackage{fix-cm}

\documentclass[twocolumn,final,natbib]{svjour3}          

\smartqed  
\usepackage{graphicx}
\usepackage{stfloats}

\graphicspath{{./Figures/}}
\usepackage{float}
\usepackage{enumitem}

\usepackage[colorlinks,pdfpagelabels,bookmarksopen,bookmarksnumbered,citecolor=blue,urlcolor=blue,linkcolor=blue]{hyperref}
\setcitestyle{numbers,square}

\usepackage[utf8x]{inputenc} 

\usepackage{mathtools}
\usepackage{amsmath}
\usepackage{amssymb}
\usepackage{amsbsy}
\usepackage{mathrsfs}
\usepackage{bm}
\usepackage{relsize}

\usepackage[flushmargin]{footmisc} 

\usepackage{empheq}

\usepackage[title]{appendix}

\usepackage[misc]{ifsym}

\usepackage{listings}
\usepackage[table]{xcolor}
\definecolor{codegreen}{rgb}{0,0.6,0}
\definecolor{codegray}{rgb}{0.1,0.1,0.1}
\definecolor{codepurple}{rgb}{0.58,0,0.82}
\definecolor{backcolour}{rgb}{1,1,1}
\lstdefinestyle{mystyle}{
	language=Matlab,
    backgroundcolor=\color{backcolour},   
    commentstyle=\color{codegreen},
    keywordstyle=\color{blue},
    deletekeywords={beta},
    morekeywords={switch,case},
    numberstyle=\tiny\color{codegray},
    stringstyle=\color{codepurple},
    basicstyle=\ttfamily\footnotesize,
    breakatwhitespace=false,         
    breaklines=true,                 
    captionpos=b,                    
    keepspaces=true,                 
    numbers=left,                    
    numbersep=1pt,                  
    showspaces=false,                
    showstringspaces=false,
    showtabs=false,                  
    tabsize=1,
    escapeinside={\%*}{*)}
}
\usepackage{cuted}

\usepackage{supertabular}  
\usepackage{array}
\newcolumntype{C}[1]{>{\centering\arraybackslash$}p{#1}<{$}} 

\usepackage{empheq}

\definecolor{Gray}{gray}{0.95}
\definecolor{LightGray}{gray}{0.9}
\definecolor{LighterGray}{gray}{0.98}

\usepackage{tikz}
\usetikzlibrary{shapes,arrows,chains}

\newcommand{\bx}{{\bf x}}
\newcommand{\bbx}{{(\bf x)}}
\newcommand{\bxhat}{{\hat{\bf x}}}
\newcommand{\bbxhat}{{(\hat{\bf x})}}
\newcommand{\bbxchi}{{({\bf x},\chi)}}
\newcommand{\bbxhatchi}{{(\hat{\bf x},\chi)}}

\newcommand{\bchi}{{(\chi)}}
\newcommand{\bchit}{{(\chi,t)}}
\newcommand{\domega}{{ \, d\Omega }}
\newcommand{\dgamma}{{ \, d\Gamma }}
\newcommand{\mcolon}{{ \, , }}
\newcommand{\mdot}{{ \, . }}

\newcommand{\codename}{{UNVARTOP}}
\newcommand{\codenamespc}{{UNVARTOP }}

\newcommand{\Kappalarge}{{\Large{\pmb \kappa}}}
\newcommand{\hS}{{r}}

\journalname{Structural and Multidisciplinary Optimization}

\journalname{ }
\def\makeheadbox{\relax}
\date{ }

\begin{document}

\title{Topology Optimization using the UNsmooth VARiational Topology OPtimization (\codename) method: an educational implementation in Matlab}
\subtitle{EDUCATIONAL ARTICLE}

\titlerunning{Topology Optimization using the UNsmooth VARiational Topology OPtimization (\codename) method}        

\author{Daniel$\ $Yago$^{1,2}$         \and
       Juan$\ $Cante$^{1,2}$           \and
       Oriol$\ $Lloberas-Valls$^{2,3}$  \and
       Javier$\ $Oliver$^{2,3}$
}

\authorrunning{Daniel Yago         \and
       		   Juan Cante           \and
       		   Oriol Lloberas-Valls  \and
       		   Javier Oliver} 

\institute{\begin{itemize}
           \item[\Letter] J. Oliver \at
                      	  \email{oliver@cimne.upc.edu} \\
           \item[1] Escola Superior d'Enginyeries Industrial, Aeroespacial i Audiovisual de Terrassa (ESEIAAT)\\
                         Technical University of Catalonia (UPC/Barcelona Tech), Campus Terrassa UPC, c/ Colom 11, 08222 Terrassa, Spain
           \item[2]      Centre Internacional de M\`{e}todes Num\`{e}rics en Enginyeria (CIMNE)\\ 
                         Campus Nord UPC, M\`{o}dul C-1 101, c/ Jordi Girona 1-3, 08034 Barcelona, Spain
           \item[3]     E.T.S d'Enginyers de Camins, Canals i Ports de Barcelona (ETSECCPB)\\
                         Technical University of Catalonia (UPC/Barcelona Tech), Campus Nord UPC, M\`{o}dul C-1, c/ Jordi Girona 1-3, 08034 Barcelona, Spain                   
           \end{itemize}	  	             	                  
}


\maketitle

\begin{abstract}
This paper presents an efficient and comprehensive MATLAB code to solve two-dimensional structural topology optimization problems, including minimum mean compliance, compliant mechanism synthesis and multi-load compliance problems. The Unsmooth Variational Topology Optimization (\codename) meth\-od, developed by \citet{Oliver2019}, is used in the topology optimization code, based on the finite element method (FEM), to compute the sensitivity and update the topology. The paper also includes instructions to improve the \emph{bisection algorithm}, modify the computation of the Lagrangian multiplier by using an Augmented Lagrangian to impose the constraint, implement heat conduction problems and extend the code to three-dimensional topology optimization problems.
The code, intended for students and newcomers in topology optimization, is included as an appendix (Appendix \ref{app_matlab_code}) and it can be downloaded from \href{https://github.com/DanielYago/UNVARTOP}{https://github.com/DanielYago} together with supplementary material.
 
\end{abstract}

\section{Introduction}
\label{sec_intro}

	The dissemination of the Matlab code, included in this paper, is intended for education purposes, in order to provide students and those new to the field with the theoretical basis for topology optimization of structural problems as well as to familiarize a wider audience with the new technique. This article is inspired by similar ones (e.g. \cite{Sigmund2001} and \cite{Andreassen2010}) which presented a Matlab implementation and possible extensions of other topology optimization approaches for structural problems.
	
	A wide variety of topology optimization approaches and the corresponding Matlab implementations can be found in the literature, including the \emph{Solid Isotropic Material with Penalization} (SIMP) method (\cite{Bendsoe1989,Bendsoe2004} and \cite{Sigmund2001}), the \emph{Bidirectional Evolutionary Structural Optimization} (BESO) method (\cite{Xie1997,Yang1999} and \cite{Zuo2015}), the \emph{Level-set} method using a shape derivative (\cite{Allaire2002,Allaire2004,Wang2004} and \cite{Challis2009,MichaelYu2004}), the parameterized Level-set method using Radial basis functions (\cite{Wang2006,Wang2007} and \cite{Wei2018}), the \emph{Topology Derivative} method (\cite{Sokolowski1999,Novotny2003} and \cite{Suresh2010}) and the \emph{Phase-field} approaches (\cite{Takezawa2010,Wang2004,Yamada2010} and \cite{Otomori2014}), among others. Along years, researchers have adapted or combined some features of these techniques to propose alternative approaches. Nevertheless, some limitations remain in any of them. 

	The \emph{Unsmooth Variational Topology Optimization} approach, first developed by \citet{Oliver2019}, appears to be an alternative to other well-established approaches due to the mathematical simplicity and robustness of the present method. So far, the \codenamespc approach has been applied in a wide range of linear applications, including static structural \cite{Oliver2019} and steady-state thermal applications \cite{Yago2020}, considering the volume constraint as a single constraint equation, with promising results, essentially in terms of computational cost. 

	The domain, in the present approach, is implicitly represented through a 0-level-set function \cite{Osher1988}, using the so-called \emph{discrimination function} $\psi$, to define a discrete \emph{characteristic function}, $\chi$, at each point of the domain, $\bx$. This variable, used as design variable, is related to the \emph{discrimination function} with the Heaviside function by $\chi\bbx={\cal{H}}(\psi\bbx)$, defining, thus, a \emph{black-and-white design}, i.e. a binary configuration with two domains: a void and a material domain. This definition is in contrast to that used by density-based methods, such as SIMP method, where the relative density, $\rho_e$, in each element is used as design variable see \citet{Bendsoe2004}. In addition, this change in the design variable, typically from \emph{Level-set methods}, allows smooth representation of the topology (void and material domains) and the corresponding boundary using the 0-level iso-surface of the \emph{discrimination function}. 
	
	The \emph{black-and-white design} is relaxed via the \emph{ersatz material approach} to a bi-material setting, where the void material is replaced with a soft material, as proposed by \citet{Allaire1997}. Despite this relaxation, the discrete nature of the \emph{characteristic function} is maintained. However, this is not true for density-based methods\footnote{The density-based methods appear also in literature as Variable Density Methods and they should be understood as synonyms throughout the text.}, which have to be relaxed via a power-law interpolation function to intermediate values (i.e. between void and solid), leading thus to the SIMP method, in order to avoid the ill-conditioning of the topology optimization problem obtaining then blurry interfaces with semi-dense elements, as stated in \citet{Sigmund1998}. 
	
	The aim of a topology optimization must be defined by means of a cost function, which will be minimized. For each specific cost function, a sensitivity evaluating the variation of it to topological perturbations must be derived. This derivation may be mathematically challenging for some topology optimization approaches. For example, the \emph{Topology Derivative} method requires heavy analytical derivation methods, dependent on the type of the topology optimization problem and the considered material in the optimization \cite{Giusti2008,Novotny2003}. However, in the current method, a consistent \emph{relaxed topological derivative} is formulated within the \emph{ersatz material approach}, and evaluated as a directional derivative of the cost function. Additionally, it can be interpreted as an approximation of the exact topological derivative, used in \emph{Topology Derivative} method, resulting in a simpler and less time-consuming derivation.
	
	Apart from the problem setting and the cost function, the procedure of updating the design variable is a crucial feature of each approach. Most of the topology optimization methods, that uses a \emph{level-set} function to define the topology layout at each iteration, update the design variable via a Hamilton-Jacobi equation using an appropriate velocity at boundaries (in terms of the precomputed sensitivity) \cite{Allaire2002,Wang2004}. Despite using an equivalent level-set function (\emph{discrimination function}), the topology is not updated neither via a Hamilton-Jacobi \cite{Wang2007,Allaire2004,Yamada2010} nor a Reaction-Diffusion \cite{Otomori2014} equations, but it is updated via the solution of a \emph{fixed-point, non-linear, closed-form algebraic system}. The fulfillment of the volume constraint is ensured within the closed-form solution by means of a \emph{Lagrange multiplier}, similar to the one used with Optimality Criteria (OC) in SIMP methods, computed through an efficient bisection algorithm. 
	
	Almost every technique require some kind of filtering in order to avoid or at least mitigate the inherent ill-posedness of the topology optimization problem \cite{Sigmund1998}. Through this filtering, the lack of mesh-independency is overcome. Density-based methods resort to density or sensitivity filtering, extensively used in density-based approaches. Nevertheless, alternative filters have been formulated in the last two decades. For instance, projection methods \cite{Guest2004} or a Helmholtz filter \cite{Lazarov2010} are also used for this purpose. This last filter, so called the \emph{Laplacian regularization} \cite{Patane2009,Vliet1989} is applied to the \emph{discrimination function} to control the filament width. A similar approach is used by \citet{Yamada2010} to control the complexity of the optimal design.
	
	Finally, the last key feature is related with the volume constraint and how the requested volume percentage is achieve. An incremental time-advancing scheme is adopted in the present methodology for the volume percentage, as a control parameter, obtaining, then, intermediate converged, optimal topologies. The optimization procedure starts from a domain fully filled with stiff material. Then, the topology optimization for a given small volume percentage is performed, obtaining a converged, optimal topology. Subsequently, the volume percentage (\emph{pseudo-time} in the algorithm) is increased and the new optimal topology is found. This procedure is repeated until the desired volume fraction is achieved, similar to the Pareto frontier-optimal tracing approach proposed by \citet{Suresh2010}. Although this implementation is not unique of the current approach, it differs from SIMP and Level-set based methods, since they directly seek the optimal topology for the requested volume fraction. Similar iterative schemes can be found in ESO/BESO approaches, where the volume fraction is incremented at each iteration until the final volume is achieved. However, optimal conditions are not fulfilled at these intermediate volumes.
	
	Thanks to this set of features, the methodology proposed in this manuscript presents a lower computational cost, around 5 times, when it is compared with other methods, e.g. a Level-set method with the RTD, while obtaining very similar results, as reported in \citet{Oliver2019} and \citet{Yago2020}. In addition, intermediate converged optimal topologies are obtained for different volume values at not additional computational cost, allowing further decisions once the topology optimization optimization has finalized.
		
	The remainder of the paper is organized as follows. The unsmooth variational topology optimization approach is briefly described in section \ref{sec_theory_approach} along with the particularities for minimum mean compliance, multi-load compliance and compliant mechanisms problems. In section \ref{sec_matlab_impl}, the code implementation of the present methodology, provided in Appendix \ref{app_matlab_code}, is discussed in detail. Several numerical examples are addressed in section \ref{sec_num_examples} to show the potential in the three optimization problems. Additionally, in section \ref{sec_extension}, possible extensions and enhancements of the code are discussed. Finally, section \ref{sec_conclusions} concludes with some final remarks.

\section{Problem formulation}
\label{sec_theory_approach}

	\subsection{Unsmooth variational topology optimization}\label{sec_thry_formulation}
		
\begin{figure}[pb]
		\includegraphics[width=7cm]{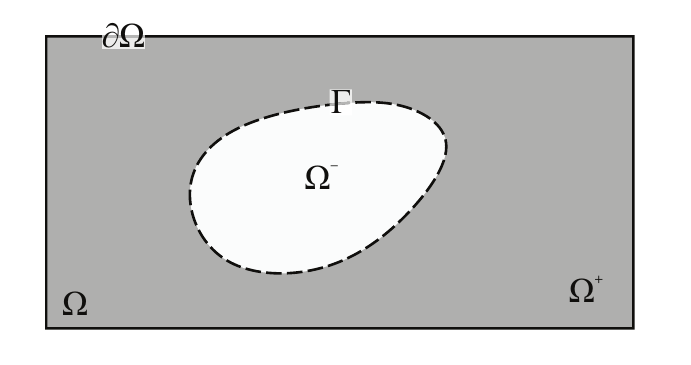}
		\caption{Representation of the fixed design domain $\Omega$.}
		\label{fig_domain_def}
	\end{figure}

	Let us define a fixed rectangular design domain, $\Omega\subset\mathbb{R}^2$, composed by two smooth subdomains, $\Omega^+$ and $\Omega^-$, as depicted in Figure \ref{fig_domain_def}. These two domains, made respectively of solid and void materials, are defined via the nonsmooth \emph{characteristic function}, $\chi\bbx:\Omega\rightarrow\{0,1\}$, as
	\begin{equation} \label{eq_characteristic_function}
		\left\{
		\begin{split}
			&\Omega^{+}\coloneqq\{\bx\in \Omega \;/\; \chi\bbx=1\} \\
			&\Omega^{-}\coloneqq\{\bx\in \Omega \;/\; \chi\bbx=0\} 
		\end{split}
		\right. \mdot
	\end{equation}
	
	\begin{figure}[pt]
		\includegraphics[width=8cm]{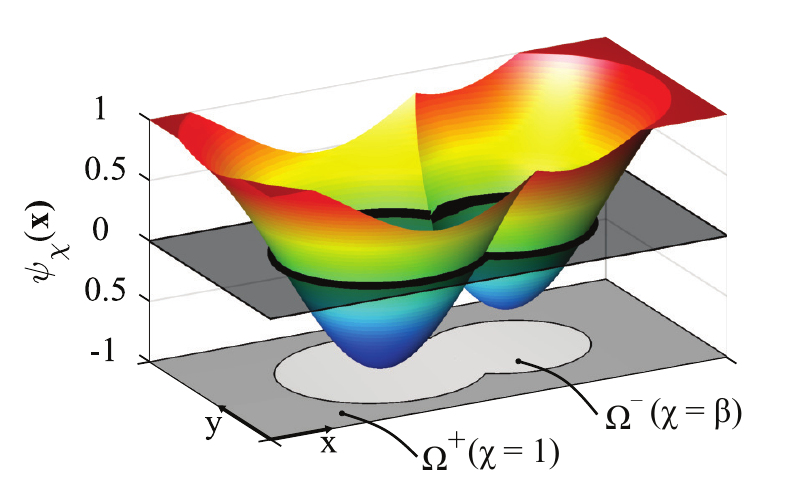}
		\caption{Topology representation in terms of the \emph{discrimination function}, $\psi$.}
		\label{fig_domain_psi_def}
	\end{figure}
	
	The topology layout can also be implicitly represented by the smooth \emph{discrimination function},  $\psi\bbx:\Omega\rightarrow\mathbb{R}$, $\psi \in H^1({\Omega})$, (see Figure \ref{fig_domain_psi_def}) defined as
	\begin{equation} \label{eq_domain_splitting}
		\left\{
		\begin{split}
			&\Omega^{+}\coloneqq\{\bx\in \Omega \;/\; \psi\bbx>0\} \\
			&\Omega^{-}\coloneqq\{\bx\in \Omega \;/\; \psi\bbx<0\} 
		\end{split}
		\right. \mdot
	\end{equation}
	In addition, the \emph{characteristic function}, $\chi_{\psi}\bbx : \Omega \rightarrow \{0,1\}$, can be expressed in terms of the \emph{discrimination function} by
	\begin{equation} \label{eq_heaviside_level_set}
		\chi_{\psi}\bbx={\cal H}(\psi\bbx) \mcolon
	\end{equation}
	where ${\cal H}(\cdot)$ stands for the Heaviside function evaluated at {$(\cdot)$}. The \emph{characteristic function}, used as the design variable, is now relaxed to $\chi_{\psi}\bbx : \Omega \rightarrow \{\beta,1\}$, where the void material is replaced with a soft material with low stiffness (ersatz material approach), with $\beta$ being the \emph{relaxation factor}.
	
	The topology optimization goal is to minimize a cost function $\mathcal{J}(\chi)$ subjected to one constraint, typically the volume, and governed by the state equations. The classic mathematical formulation of the corresponding topology optimization problem can be expressed as
	\begin{equation} \label{eq_minimization_restricted}
		\left[\begin{split}
			&\underset{\chi\in{\mathscr{U}}_{ad}} {\operatorname{min}}\quad{\cal J}\left( \chi\right)\equiv\int_{\Omega}{j(\chi,\bx)}\domega &\quad(a) \\
			&\text{subject to:} \\
			&\hspace{1.2cm}{\cal C}(\chi)\equiv\int_{\Omega}{c(\chi,\bx)\domega}=0 &\quad(b) \\
			&\text{governed by:} \\
			&\hspace{1.2cm}{\textstyle Equilibrium \; equation } &\quad(c) 
		\end{split}\right. \mcolon
	\end{equation}
	where ${\mathscr{U}}_{ad}$ stands for the set of admissible solutions for $\chi$ and ${\cal C}\bchi$ represents the constraint functional (e.g. the volume constraint).
		
	Following \citet{Oliver2019}, the \emph{Relaxed Topological Derivative} (RTD), specific characteristic of the proposed approach, evaluated as 
	\begin{equation} \label{eq_RTD_coste}
		\dfrac{\delta {\cal J}(\chi)}{\delta \chi}\bbxhat=
		\left[\dfrac{\partial j({\chi},\bx)}{\partial {\chi}}\right]_{\bx=\bxhat}\Delta \chi\bbxhat \mcolon
	\end{equation}
	measures the sensitivity of the functional (\ref{eq_minimization_restricted})-a, in terms of the classical Fréchet derivative $\frac{\partial(\cdot)}{\partial\chi}\bbxhat$ of the integral kernel, when a material exchange is made at point $\bxhat$. The term ${\Delta \chi}\bbxhat$, denoted as the \emph{exchange function}, corresponds to the signed variation of $\chi\bbxhat$, due to that material exchange, i.e. 
	\begin{equation} \label{eq_alpha}
		{\Delta\chi}\bbxhat=
		\left\{
		\begin{split}
			-(&1-\beta)<0  \;\;\;\;\textit{for} \;\;\bxhat\in\Omega^{+}  \\ 
			 (&1-\beta)>0 \;\;\;\;\textit{for} \;\;\bxhat\in\Omega^{-}
		\end{split} 
		\right. \mdot
	\end{equation}
	Notice that the RTD of equation (\ref{eq_RTD_coste}) will depend on each specific cost function, as detailed in sections \ref{sec_thry_compliance} to \ref{sec_thry_mechanism}. Mimicking equation (\ref{eq_RTD_coste}), the RTD of the volume constraint ((\ref{eq_minimization_restricted})-b) is computed as
	\begin{equation} \label{eq_RTD_constraint}
		\dfrac{\delta {\cal C}(\chi,t)}{\delta \chi}\bbxhat=
		\left[\dfrac{\partial c({\chi},\bx)}{\partial {\chi}}\right]_{\bx=\bxhat}\Delta \chi\bbxhat=\frac{1}{\vert\Omega\vert}\text{sgn}(\Delta\chi\bbxhat) \mcolon
	\end{equation}
	where ${\mathcal C}(\chi,t)\coloneqq t-\frac{\vert\Omega^-\vert(\chi)}{\vert\Omega\vert}=0$ and $\vert\Omega^-\vert(\chi)=\int_{\Omega} \frac{1-\chi}{1-\beta} \domega$. Additionally, the $sgn(\cdot)$ corresponds to the sign function of $(\cdot)$, while the term $t\in[0,T]$ corresponds to the \emph{pseudo-time} parameter, given by the user, used in the \emph{pseudo-time-advancing strategy}. Notice that the parameter $T$ stands for the \emph{pseudo-time} corresponding to the final volume.
	
	The \emph{Lagrangian function} of the optimization problem (\ref{eq_minimization_restricted}) can be commonly expressed as
	\begin{equation}\label{eq_Lagrangian}
		{\cal L}\bchi={\cal J}\bchi+\lambda {\cal C}\bchit \mcolon
	\end{equation}
	where the \emph{constraint equation}, $\cal{C}$, multiplied with a Lagrange multiplier, $\lambda$, is added to the original cost function $\cal{J}$. The value of $\lambda$ is such that the volume constraint is fulfilled.
	
	Finally, applying the RTD to equation (\ref{eq_Lagrangian}) and considering equations (\ref{eq_RTD_coste}) and (\ref{eq_RTD_constraint}), the optimality condition of the \emph{original topology optimization problem} can be written as
	\begin{multline} \label{eq_optimal_condition}
			\dfrac{\delta {\cal L}(\chi,\lambda)}{\delta\chi}\bbxhat=\left(\dfrac{\partial j \left(\chi,\bxhat \right)}{\partial \chi}{\Delta\chi}\bbxhat
				+\lambda\,\text{sgn}(\Delta\chi\bbxhat)\right)= \\
			= - \psi\bbxhatchi = - (\xi\bbxhatchi - \lambda) \;\; \forall\bxhat\in\Omega \mcolon
	\end{multline}
	where $\psi\bbxhatchi$ corresponds to the \emph{discrimination function} and $\xi\bbxhatchi$ is termed the \emph{pseudo-energy} and must be computed for each optimization problem, thus obtaining a similar updating expression to other topology optimization techniques\footnote{Note that, since $\vert\Omega\vert$ is constant, it can be included in the Lagrange multiplier, $\lambda$, in the second term of the equation (\ref{eq_optimal_condition}).}. Compared to other techniques, the \emph{pseudo-energy} is first shifted\footnote{The shifting is applied in order to obtain positive \emph{pseudo-energy}, $\xi$, in $\Omega$ at $t=0$, thus, ensuring a converged topology for this time-step.} and normalized, yielding to the \emph{modified energy density} defined as
	\begin{equation} \label{eq_thry_shift_norm}
		\hat{\xi}\bbxhat=\dfrac{\xi\bbxhat-\chi\bbxhat\Delta_{shift}}{\Delta_{norm}} \mcolon
	\end{equation}
	where $\Delta_{shift}$ and $\Delta_{norm}$ correspond to the shifting and normalization parameters defined at the first iteration as $\operatorname{min}(\xi_0,0)$ and $\operatorname{max}(\operatorname{range}(\xi_0),\operatorname{max}(\xi_0))$, respectively. The resultant $\psi$, after replacing equation (\ref{eq_thry_shift_norm}) into (\ref{eq_optimal_condition}), is subsequently smoothed through a \emph{Laplacian regularization}, in contrast to other distance-based filters used in methods such as SIMP or ESO, in order to mitigate mesh-dependency along with controlling the minimum filament's size. The \emph{smooth discrimination function}, $\psi_\tau$, corresponds to the solution of
	\begin{equation} \label{eq_regularized_laplacean_smoothing_psi}
		\left\{
		\begin{split}
			&\psi_{\tau}-(\tau h_e)^2\Delta_\bx\psi_{\tau}=\psi& &\quad in \; \Omega\\
			&\nabla_\bx\psi_{\tau}\cdot\mathbf{n}={0}& &\quad on \; \partial\Omega
		\end{split}
		\right. \mcolon
	\end{equation}
	where, $\Delta_\bx(\bx,\cdot)$ and $\nabla_\bx(\bx,\cdot)$ are respectively the Laplacian and gradient operators, and $\mathbf{n}$ is the outwards normal to the boundary of the design domain, $\partial\Omega$. $\tau$ and $h_e$ stand for the dimensionless \emph{regularization parameter} and the typical size of the finite element mesh, respectively.
	
	The topology layout, $\chi$, is updated by means of the \emph{Cutting}\&\emph{Bisection} algorithm\footnote{The present \emph{Cutting}\&\emph{Bisection} algorithm has been so far applied to single constrained topology optimization problems subject to equality, pseudo-time evolving volume constraints. Further development is required to extend it to other constraints.}, in which the value of $\lambda$, which enforces volume constraint (equation (\ref{eq_minimization_restricted})-b), is computed. Then, a \emph{closed-form solution} of the topology optimization problem (\ref{eq_minimization_restricted}) can be written as
	\begin{equation} \label{eq_solution_psi_xi}
		\left\{
		\begin{split}
			&\psi\bbxhat\coloneqq\hat{\xi}\bbxhatchi-\lambda \quad \\
			&\chi \bbxhat={\cal H}(\psi_\tau\bbxhat) \\
			&{\cal C}(\chi(\lambda),t)= 0
		\end{split}
		\right.
		in \; \Omega \mcolon
	\end{equation}
	where $\psi_\tau\bbxhat$ corresponds to the solution of equation (\ref{eq_regularized_laplacean_smoothing_psi}) that must be applied at each iteration. Equation (\ref{eq_solution_psi_xi}) constitutes a fundamental feature of the \codenamespc method, as aforementioned in section \ref{sec_intro}. Nonetheless, the \emph{Laplacian regularization} only affects the \emph{modified energy density}, $\hat{\xi}\bbxhatchi$, since the term $\lambda$ is constant, thus leading equation (\ref{eq_solution_psi_xi}) to 
	\begin{equation} \label{eq_solution_psi_xi_2}
		\left\{
		\begin{split}
			&\psi_\tau\bbxhat\coloneqq\hat{\xi}_\tau\bbxhatchi-\lambda \quad \\
			&\chi \bbxhat={\cal H}(\psi_\tau\bbxhat) \\
			&{\cal C}(\chi(\lambda),t)= 0
		\end{split}
		\right.
		in \; \Omega \mcolon
	\end{equation}
	where $\hat{\xi}_\tau$ is the solution of equation (\ref{eq_regularized_laplacean_smoothing_psi}) for the \emph{modified energy density}. Due to this modification, the computational cost of the bisection algorithm is significantly reduced.
			
	For more details on the formulation, the reader is referred to \citet{Oliver2019} and \citet{Yago2020}, where in-depth discussions are made on each subject.
	
	\subsection{State problem}\label{sec_thry_state_eq}
	
	The governing variational problem for linear elasticity, in terms of the displacement field ($\mathbf{u}_\chi$) and the virtual displacement field ($\mathbf{w}$), can be written as
	\begin{empheq}[left=\empheqlbrack,right=\hspace{-0.2cm}]{align}
		&\text{Find the displacement field  ${\pmb u}_\chi\in{\cal{U}}(\Omega)$ such that} \notag \\
		& \hspace{0.25cm} a(\mathbf{w},\mathbf{u}_\chi) = l(\mathbf{w}) \quad \forall \mathbf{w}\in {\cal V}(\Omega) \label{eq_weak_problem2}\\
	    &\text{where} \notag \\
		&\hspace{0.25cm}a(\mathbf{w},\mathbf{u}_\chi) = \int_{\Omega}{\bm{\nabla}^S \mathbf{w}\bbx:\mathbb{C}_\chi\bbx:\bm{\nabla}^S\mathbf{u}_\chi\bbx}\domega\mcolon \label{eq_lhs_structural_problem} \\
		&\hspace{0.25cm}l(\mathbf{w}) = \int_{\partial_{\sigma}\Omega}{\mathbf{w}\bbx\cdot{\overline{\pmb{\sigma}}}\bbx}\dgamma  \notag \\
		&\hspace{1.25cm}+ \int_{\Omega}{\mathbf{w}\bbx \cdot \mathbf{b}_\chi\bbx}\domega \mcolon \label{eq_rhs_structural_problem}
	\end{empheq}
	where $\mathbb{C}_\chi$ and $\mathbf{b}_\chi$ correspond to the fourth order elastic constitutive tensor and the volumetric force, respectively. In addition, ${\overline{\pmb{\sigma}}}\bbx$ stands for the boundary tractions applied on $\partial_{\sigma}\Omega\subset\partial\Omega$, while the term $\bm{\nabla}^S(\cdot)$ corresponds to the symmetrical gradient of $(\cdot)$. Finally, the set of admissible displacement fields, ${\cal{U}}(\Omega)$, is defined as  ${\cal{U}}(\Omega) \coloneqq \left\{\mathbf{u}\bbx \; / \;  \mathbf{u}\in H^1(\Omega) , \; \mathbf{u} = \overline{\mathbf{u}} \; on \; \partial_{u}\Omega \right\}$, while the space of admissible virtual displacement fields is given by ${\cal V}(\Omega)  \coloneqq \left\{\mathbf{w}\bbx      \; / \; \mathbf{w}      \in H^1(\Omega) , \; \mathbf{w} = 0                 \; on \; \partial_{u}\Omega  \right\}$.
	
	As in any other topology optimization approach, the constitutive tensor\footnote{The constitutive tensor is governed by Hooke's law, i.e. $\pmb{\sigma}=\mathbb{C}\pmb{\varepsilon}$, with $\pmb{\varepsilon}$ being the strain tensor ($\pmb{\varepsilon} = {\bm{\nabla}^S u}_{\chi}\bbx$).}, $\mathbb{C}_\chi$,  and the volumetric force, $\mathbf{b}_\chi$, depend on the topology. Thus, they are mathematically defined in terms of the \emph{characteristic function} as follows
	\begin{empheq}[left=\empheqlbrace]{align}
		& \mathbb{C}_\chi\bbx = \chi_{k} ^{m_{k}} \bbx \overline{\mathbb{C}}\bbx\; ; \quad m_{k} > 1 \label{eq_stiff_interp}\\
		& \mathbf{b}_{\chi}\bbx = \chi_{_{b}} ^{m_{b}}\bbx \overline{\mathbf{b}}\bbx\; ; \quad \; m_{b} > 1\label{eq_volforce_interp}
	\end{empheq} 
	where $m_{(\cdot)}$ stands for the \emph{exponential factor} of property $(\cdot)$. The lower limit of the \emph{relaxed characteristic function}, $\chi_\beta$, is defined through the \emph{contrast factor}, $\alpha_{(\cdot)}$, and $m_{(\cdot)}$ by $\beta_{(\cdot)}=\alpha_{(\cdot)} ^{1/{m_{(\cdot)}}}$. Both $\overline{\mathbb{C}}$ and $\overline{\mathbf{b}}$ denote the corresponding nominal property of the stiff material.
	
	Assuming plane-stress condition, the constitutive tensor $\overline{\mathbb{C}}$ is given by
	\begin{equation} \label{eq_thry_Pstress}
		\overline{\mathbb{C}}^{Pstress}=\dfrac{E}{1-\nu^2}\left[  
		\begin{array}{*3{C{3em}}}
			1 & \nu & 0 \\
			\nu & 1 & 0  \\
			0   & 0     & \dfrac{1-\nu}{2} \\
		\end{array}
		\right] \mcolon
	\end{equation}
	with $E$ representing the Young's modulus of the \emph{stiff material} and $\nu$, the Poisson's ratio of the isotropic material. 
	
	\subsection{Finite element discretization}\label{sec_thry_finite_element}
	
	The state equation (\ref{eq_weak_problem2}) is now discretized using the standard finite element method \cite{Zienkiewicz2013,Rao2004}. The displacement field and its gradient are approximated as follows
	\begin{align} 
		&\mathbf{u}_{\chi}\bbx\equiv\mathbf{N}_{u}\bbx\hat{\pmb{u}}_\chi \label{eq_shape_disp}\\
		&{\bm{\nabla}^S u}_{\chi}\bbx\equiv\mathbf{B}\bbx\hat{\pmb{u}}_\chi \label{eq_grad_disp}
	\end{align}
	where $\mathbf{N}_{u}\bbx$ and $\mathbf{B}\bbx$ stand for the displacement, shape function matrix and the strain-displacement matrix, respectively, and $\hat{\pmb{u}}_\chi$ corresponds to the nodal displacement vector. 
	
	Introducing equations (\ref{eq_stiff_interp})-(\ref{eq_volforce_interp}) and (\ref{eq_shape_disp})-(\ref{eq_grad_disp}) into equations (\ref{eq_weak_problem2})-(\ref{eq_rhs_structural_problem}), the resultant state equation reads
	\begin{equation} \label{eq_equilibrium}
		{\mathbb K}_{\chi}\hat{\pmb{u}}_{\chi}=\mathbf{f} 
	\end{equation}
	with
	\begin{equation} \label{eq_equilibrium_stiff_force}
		\left\{
		\begin{split} 
			&{\mathbb K}_{\chi}=\int_{\Omega}\mathbf{B}^{\text{T}}\bbx\ {\mathbb{C}}_{\chi}\bbx\ \mathbf{B}\bbx \domega \\
			&\begin{split}
				\mathbf{f}=&\int_{\partial_{\sigma}\Omega}\mathbf{N_{u}}^{\text{T}}\bbx{\overline{\pmb{\sigma}}}\bbx\dgamma \\
				&+ \int_{\Omega} \mathbf{N_{u}}^{\text{T}}\bbx \mathbf{b}_{\chi}\bbx\domega 
			\end{split}
		\end{split} \right. \mcolon
	\end{equation}
	where $\mathbb{K}_{\chi}$ and $\mathbf{f}$ stand for the stiffness matrix and the external forces vector, respectively. The element stiffness matrix and the volumetric term of the force vector are numerically integrated inside each element, $\Omega_e$, employing several quadrature points. Subsequently, these terms are assembled to obtain the global stiffness matrix and force vector.
	
	\subsection{Algorithm}\label{sec_thry_algorithm}
	
	The flowchart of the algorithm used to obtain the optimal topology layouts in terms of the \emph{characteristic function}, $\chi$, is illustrated in Figure \ref{fig_algorithm}. 
	
	\colorlet{lcnorm}{black}
	\begin{figure}
		\centering
		\begin{tikzpicture}[%
		    >=triangle 60,              
		    start chain=going below,    
		    node distance=5mm and 60mm, 
		    every join/.style={norm},   
		    ]
		\tikzset{
		  base/.style={draw, on chain, on grid, align=center, minimum height=1ex},
		  proc/.style={base, rectangle, text width=15em},
		  test/.style={base, diamond, aspect=2, text width=5em},
		  term/.style={proc, rounded corners},
		  coord/.style={coordinate, on chain, on grid, node distance=4mm and 35mm},
		  norm/.style={->, draw, lcnorm},
		  it/.style={font={\small\itshape}}
		}
		\node [term]       			{Start Topology optimization};
		\node [proc, join] 	(p1)  	{\hyperref[sec_code_parameters]{Preprocessing and data initialization} ($t=t_0$) [2-58]}; 
		\node [proc, join] 	(p2)  	{Increase \emph{pseudo-time} ($t_{n+1}=t_{n}+\Delta t _{n+1}$) [60]};
		\node [proc, join] 	(p3)  	{\hyperref[sec_code_FEM]{Solve equilibrium equations (FEM)} and \hyperref[sec_code_sensitivity]{compute sensitivity} [66-79]};
		\node [proc, join] 	(p6)  	{\hyperlink{sentcodecostfunction}{Compute cost function} [69]}; 
		\node [proc, join] 	(p4)  	{\hyperref[sec_lapl_reg_code]{Apply Laplacian regularization} [80-86]}; 
		\node [proc, join] 	(p5)  	{\hyperref[sec_code_bisection]{Compute Lagrangian multiplier} [88]};
		\node [proc, join] 	(p7)  	{\hyperref[sec_code_update_top]{Update topology} ($\psi$ and $\chi$) [88]}; 
		\node [test, join] 	(t1) 	{\hyperref[sec_code_convergence]{Convergence}? [65,90,94]}; 
		\node [proc] 		(p8)    {\hyperref[sec_code_plot_cost]{Optimal topology layout} [100-103]};
		\node [test, join] 	(t2) 	{Last time-step? [59]};
		\node [term] 		(p9)    {\hyperref[sec_code_plot_gui]{Post-processing} [109] and Exit}; 
		\node [coord, right=of p3] 	(c1)  {}; 
		\node [coord, right=of t1] 	(c2)  {}; 
		\node [coord, left=of t2] 	(c3)  {}; 
		\node [coord, left=of p2] 	(c4)  {}; 
		\path (t1.south) to node [near start, xshift=1em] {$yes$} (p8);
		  \draw [*->,lcnorm] (t1.south) -- (p8);
		\path (t2.south) to node [near start, xshift=1em] {$yes$} (p9); 
		  \draw [*->,lcnorm] (t2.south) -- (p9);
		\path (t2.west) to node [yshift=-1em] {$no$} (c3); 
		  \draw [*->,lcnorm] (t2.west) -- (c3) |- (p2);
		\path (t1.east) to node [yshift=1em] {$no$} (c2); 
		  \draw [*->,lcnorm] (t1.east) -- (c2) |- (p3);
		\end{tikzpicture}
		\caption{The flowchart for the unsmooth variational topology optimization algorithm with the corresponding code lines in brackets.}
		\label{fig_algorithm}
	\end{figure}
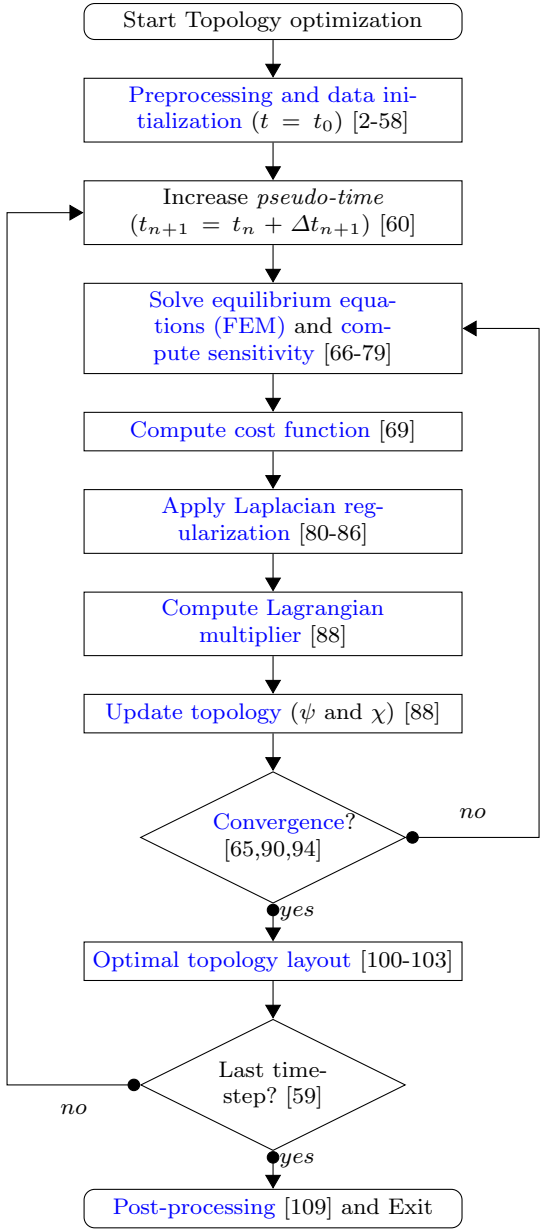
	
	The algorithm is based on a two-steps procedure: 1) data initialization and FE analysis pre-processing, e.g. mesh generation, creation of figures, computation of element FE matrices, assembly of Laplacian regularization matrix, along others, and 2) a topology optimization loop over time-steps. For each step, the state equation (\ref{eq_equilibrium}) is solved to obtain the displacement vector, and the corresponding sensitivities are computed (equations (\ref{eq_RTD_coste}) and (\ref{eq_RTD_constraint})), obtaining then the \emph{pseudo-energy}, $\xi$, dependent on each topology optimization problem defined in subsequent sections, and the corresponding \emph{modified energy density}, $\hat{\xi}$ (equation (\ref{eq_thry_shift_norm})). The cost function is then computed via equation (\ref{eq_minimization_restricted})-a using the previously computed displacement vector. Then, the \emph{Laplacian regularization} is applied to $\hat{\xi}$ (equation (\ref{eq_regularized_laplacean_smoothing_psi})) while the Lagrange multiplier is obtained by means of a \emph{bisection algorithm} (equation (\ref{eq_solution_psi_xi_2})), thus obtaining the new optimal topology (in terms of the \emph{discrimination function} $\psi$ and the corresponding \emph{characteristic function} $\chi$). If tolerances are fulfilled\footnote{The L2-norm of the \emph{characteristic function} and the L$\infty$-norm of the Lagrange multiplier are checked.}, the topology is considered as converged and then the \emph{pseudo-time}, $t$, is increased. Otherwise, an iteration is carried out with the new topology.
		
	\subsection{Mean compliance}\label{sec_thry_compliance}
	
	The main goal of the minimum mean compliance problems is to seek the \emph{optimal topology layout}, in terms of the \emph{characteristic function}, $\chi$, that maximizes the global stiffness of the structure given specific boundary conditions. That is, the external work produced by applied forces is minimized. The objective function is written as
	\begin{equation} \label{eq_compliance_cost_function}
		\begin{split}
			{\cal J}(\mathbf{u}_\chi)&\equiv l(\mathbf{u}_{\chi})\equiv a_{\chi}(\mathbf{u}_{\chi},\mathbf{u}_{\chi}) \equiv \\
				&\equiv 2  \int_{\Omega} \frac{1}{2} \bm{\nabla}^S\mathbf{u}_{\chi} :{\mathbb{C}}_{\chi}:\bm{\nabla}^S\mathbf{u}_{\chi}\domega = 2 \int_\Omega {\cal U_\chi}\domega \mcolon
		\end{split}
	\end{equation}
	where ${\cal U}_\chi$ can be identified as the \emph{actual strain energy density} (${\cal U}_{\chi}=\frac{1}{2}\bm{\nabla}^S\mathbf{u}_{\chi}:{\mathbb{C}}_{\chi}:\bm{\nabla}^S\mathbf{u}_{\chi}$), and $a_{\chi}(\mathbf{u}_{\chi},\mathbf{u}_{\chi})$ and $l(\mathbf{u}_{\chi})$ are the bilinear forms of the elastic problem (\ref{eq_weak_problem2}) for $\mathbf{w}=\mathbf{u}_\chi$.
	
	Considering equations (\ref{eq_equilibrium}) and (\ref{eq_compliance_cost_function}), the corresponding finite element discretization counterpart of problem (\ref{eq_minimization_restricted}) reads
	\begin{equation}\label{eq_topopt_prob_compliance}
		\left[
		\begin{split}
			&\underset{\chi\in{\mathscr{U}}_{ad}}{\operatorname{min}}\ {\cal J}^{(h_e)}(\mathbf{u}_{\chi}(t))\equiv \mathbf{f}^{\text{T}}\hat{\pmb{u}}_{\chi}(t)\quad&(a)  \\
			&\text{subject to:} \\
			&\hspace{0.25cm}{\mathcal C}(\chi,t)\coloneqq t-\dfrac{\vert\Omega^-\vert(\chi)}{\vert\Omega\vert}=0\;;\quad t\in[0,1]   &(b) \\
			&\text{governed by:}  \\
			&\hspace{0.25cm}{\mathbb K}_{\chi}\hat{\pmb{u}}_{\chi}=\mathbf{f} &(c)
		\end{split} \right. \mcolon
	\end{equation}
	where $\mathbf{f}^{\text{T}}\hat{\pmb{u}}_{\chi}$ denotes the structural compliance. 
	
	According to \citet{Oliver2019}, the \emph{relaxed topological derivative} with respect to $\chi\bbx$, using the adjoint method\footnote{The adjoint method is used to avoid explicitly compute the sensitivities of the displacements. The minimum mean compliance problem is self-adjoint.}, is defined as
	\begin{equation} \label{eq_topopt_derivative_compliance}
		\dfrac{\delta\overline{\cal J}^{(h_e)}(\chi)}{\delta\chi}\bbxhat = \left[2 \dfrac{\delta\mathbf{f}^{\text{T}} _{\chi}}{\delta{\chi}}\bbx\hat{\pmb{u}}_{\chi}- \hat{\pmb{u}}_{\chi}^{\text{T}}\dfrac{\delta {\mathbb K}_{\chi}}{\delta\chi}\bbx\hat{\pmb{u}}_{\chi}\right]_{\bx=\bxhat} \mdot
	\end{equation}
	
	Assuming that no volumetric forces are applied on the domain and substituting the definition of the relaxed topological derivative of each term (\ref{eq_RTD_coste}), equation (\ref{eq_topopt_derivative_compliance}) can be expressed as
	\begin{equation} \label{eq_topopt_derivative_energy_compliance}
		\begin{split}
			\dfrac{\delta{\overline{\cal J}^{(h_e)}(\mathbf{u}_{\chi})}}{\delta\chi}\bbxhat&=-2m_{k} \left(\chi_{k}\bbxhat\right) ^{m_{k}-1}{\overline{\cal U}}\bbxhat{{\Delta\chi}_{k}\bbxhat } \mcolon
		\end{split}
	\end{equation}
	where the \emph{nominal energy density}, $\overline{\cal U}\bbxhat$, is given by
	\begin{equation}
		\overline{\cal U}\bbxhat=\dfrac{1}{2}\left( \bm{\nabla}^S\mathbf{u}_{\chi} : {\overline{\mathbb{C}}} : \bm{\nabla}^S\mathbf{u}_{\chi}\right)\bbxhat \ge 0 \mdot
	\end{equation}
	
	Finally, comparing equation (\ref{eq_topopt_derivative_energy_compliance}) with equation (\ref{eq_optimal_condition}), the \emph{pseudo-energy}, $\xi\bbxhatchi$, of topology problem (\ref{eq_topopt_prob_compliance}) reads
	\begin{equation} \label{eq_topopt_xi_compliance}
		\xi\bbxhatchi = 2m_{k} \left(\chi_{k}\bbxhat\right) ^{m_{k}-1}{\overline{\cal U}}\bbxhat{{\Delta\chi}_{k}\bbxhat } \mcolon
	\end{equation}
	which must be then modified as detailed in equation (\ref{eq_thry_shift_norm}). Discretizing the terms in equation (\ref{eq_topopt_xi_compliance}), and after some mathematical manipulations, it can be numerically computed as
	\begin{equation} \label{eq_topopt_xi_code_compliance}
		\xi\bbxhatchi = \gamma_1 {\mathbf{\hat{u}}}\bbxhat^{\text{T}} \left[{\mathbf{B}}\bbxhat^{\text{T}} \overline{\mathbb{C}} {\mathbf{B}}\bbxhat \right] {\mathbf{\hat{u}}}\bbxhat \mcolon
	\end{equation}
	with $\gamma_1=2m_{k} \left(\chi_{k}\bbxhat\right) ^{m_{k}-1} {\Delta\chi}_{k}\bbxhat$. 
	
	\subsection{Multi-load mean compliance} \label{sec_thry_multicompliance}
		
	Multi-load compliance problems are considered a specific case of minimum compliance problems (see section \ref{sec_thry_compliance}), in which a set of elastic problems with different loading conditions are solved independently. The objective function (\ref{eq_compliance_cost_function}) is replaced with the weighted average sum of all the cases, i.e.
	\begin{equation} \label{eq_multi_cost_function}
		\begin{split}
			{\cal J}(\mathbf{u}_\chi)&\equiv \sum_{i=1}^{n_l}  l\left(\mathbf{u}_{\chi}^{(i)}\right)\equiv\\
			&\equiv \sum_{i=1}^{n_l}  \int_{\Omega} \bm{\nabla}^S\mathbf{u}_{\chi}^{(i)} :{\mathbb{C}}_{\chi}:\bm{\nabla}^S\mathbf{u}_{\chi}^{(i)}\domega = \\
			& = \sum_{i=1}^{n_l} 2 \int_\Omega {{\cal U}_\chi ^{(i)} }\domega \mcolon
		\end{split}
	\end{equation}
	where $n_l$ stands for the number of loading states and ${{\cal U}_\chi ^{(i)}}$ corresponds to the \emph{actual energy density} of the $i$-th loading state. Then, according to this new definition, equation (\ref{eq_topopt_prob_compliance}) is rewritten as
	\begin{equation}\label{eq_topopt_prob_multi}
		\left[
		\begin{split}
			&\underset{\chi\in{\mathscr{U}}_{ad}}{\operatorname{min}}\ {\cal J}^{(h_e)}(\mathbf{u}_{\chi}(t))\equiv \sum_{i=1}^{n_l} \mathbf{f}^{(i) \text{T}}\hat{\pmb{u}}_{\chi}^{(i)}(t)\quad&(a)  \\
			&\text{subject to:} \\
			&\hspace{0.25cm}{\mathcal C}(\chi,t)\coloneqq t-\dfrac{\vert\Omega^-\vert(\chi)}{\vert\Omega\vert}=0\;;\quad t\in[0,1]   &(b) \\
			&\text{governed by:}  \\
			&\hspace{0.25cm}{\mathbb K}_{\chi}\hat{\pmb{u}}_{\chi}^{(i)}=\mathbf{f}^{(i)} \quad \forall i \in [1,n_l] &(c)
		\end{split} \right. \mdot
	\end{equation}
	
	Equations (\ref{eq_topopt_derivative_compliance}) to (\ref{eq_topopt_xi_compliance}) are consequently modified to account multiple loading cases, leading to 
	\begin{equation} \label{eq_topopt_xi_code_multi}
		\xi\bbxhatchi = \gamma_1 \sum_{i=1}^{n_l} {\mathbf{\hat{u}}}\bbxhat^{(i) \text{T}} \left[{\mathbf{B}}\bbxhat^{\text{T}} \overline{\mathbb{C}} {\mathbf{B}}\bbxhat \right] {\mathbf{\hat{u}}}^{(i)}\bbxhat \mdot
	\end{equation}
		
	Bear in mind that the optimal topology layout will considerably differ from the single minimum compliance problem with all the loads applied at the same time.	Multi-load optimization problems are employed to find a trade-off between optimal topologies for each loading state.
	
	\subsection{Compliant mechanisms}\label{sec_thry_mechanism}
	
	Compliant mechanisms are flexible structures that transfer an action (force or displacement) at the \emph{input port} to the \emph{output port}, obtaining a desired force or displacement at that port. The objective function, ${\cal J}$, can be expressed in terms of the displacement at the output port, when maximum displacement is sought, as
	\begin{equation} \label{eq_complmech_cost_function}
		{\cal J}(\mathbf{u}_\chi)\equiv {\mathbf{1}} ^{\text{T}} {\mathbf{\hat{u}}}_\chi \mcolon
	\end{equation}
	where ${\mathbf{1}}$ represents a dummy constant force vector applied only on the \emph{output port} at the desired direction. Additional springs, denoted by $K_{in}$ and $K_{out}$, must be considered in the \emph{input} and \emph{output ports}, respectively.
	
	In the context of finite element discretization, like in equation (\ref{eq_topopt_prob_compliance}), the \emph{topology optimization problem} (\ref{eq_minimization_restricted}) can be expressed as
	\begin{equation}\label{eq_topopt_prob_complmech}
		\left[
		\begin{split}
			&\underset{\chi\in{\mathscr{U}}_{ad}}{\operatorname{min}}\ {\cal J}^{(h_e)}(\mathbf{u}_{\chi}(t))\equiv -\mathbf{1}^{\text{T}}\hat{\pmb{u}}_{\chi}(t)\quad&(a)  \\
			&\text{subject to:} \\
			&\hspace{0.25cm}{\mathcal C}(\chi,t)\coloneqq t-\dfrac{\vert\Omega^-\vert(\chi)}{\vert\Omega\vert}=0\;;\quad t\in[0,1]   &(b) \\
			&\text{governed by:}  \\
			&\hspace{0.25cm}{\mathbb K}_{\chi}\hat{\pmb{u}}_{\chi}=\mathbf{f} &(c)
		\end{split} \right. \mcolon
	\end{equation}
	where the cost function (\ref{eq_complmech_cost_function}) has been defined as a minimization problem by changing its sign.
	
	Contrary to the problem of minimal compliance (section \ref{sec_thry_compliance}), the compliant mechanism problem is not self-adjoint. Thus, an \emph{auxiliary state problem} must be solved in addition to the \emph{original state problem} (\ref{eq_equilibrium}). Both systems present the same stiffness matrix ${\mathbb K}_{\chi}$ but different actions and solutions $\hat{\pmb{u}}_{\chi}^{(1)}$ and $\hat{\pmb{u}}_{\chi}^{(2)}$, respectively, defined as
	\begin{equation} \label{eq_complmech_adjoint}
		\left\{
		\begin{split}
			&{\mathbb K}_{\chi}\ \hat{\pmb{{u}}}_{\chi}^{(1)}={\bf f}^{(1)}\quad& &(\text{system  I})\\
			&{\mathbb K}_{\chi}\ \hat{\pmb{{u}}}_{\chi}^{(2)}=\mathbf{1}\quad& &(\text{system II})\\
		\end{split}
		\right.
	\end{equation}
	
	Following \citet{Oliver2019}, the \emph{relaxed topological derivative} of the optimization problem (\ref{eq_topopt_prob_complmech}), once the \emph{adjoint state equation} (\ref{eq_complmech_adjoint}) has been substituted in, can be expressed as
	\begin{equation} \label{eq_topopt_derivative_complmech}
		\dfrac{\delta\overline{\cal J}^{(h_e)}(\chi)}{\delta\chi}\bbxhat = \left[\hat{\pmb{u}}_{\chi}^{(2)\ \text{T}}\dfrac{\delta {\mathbb K}_{\chi}}{\delta\chi}\bbx\hat{\pmb{u}}_{\chi}^{(1)} -\hat{\pmb{u}}_{\chi}^{(2)\ \text{T}} \dfrac{\delta\mathbf{f}^{(1)} _{\chi}}{\delta{\chi}}\bbx \right]_{\bx=\bxhat} \mdot
	\end{equation}
	
	As proceed in section \ref{sec_thry_compliance}, equation (\ref{eq_topopt_derivative_complmech}) can be simplified and expressed in terms of a \emph{pseudo-energy density}, yielding to
	\begin{equation} \label{eq_topopt_derivative_energy_complmech}
		\dfrac{\delta{\overline{\cal J}^{(h_e)}(\mathbf{u}_{\chi})}}{\delta\chi}\bbxhat=2m_{k} \left(\chi_{k}\bbxhat\right) ^{m_{k}-1}{\overline{\cal U}}_{1-2}\bbxhat{{\Delta\chi}_{k}\bbxhat } \mcolon
	\end{equation}
	when volumetric forces are neglected. The corresponding \emph{nominal pseudo-energy density} can be determined as
	\begin{equation}
		\overline{\cal U}_{1-2}\bbxhat=\dfrac{1}{2}\left( \bm{\nabla}^S\mathbf{u}_{\chi}^{(2)} : {\overline{\mathbb{C}}} : \bm{\nabla}^S\mathbf{u}_{\chi}^{(1)}\right)\bbxhat \mdot
	\end{equation}
	
	Finally, mimicking equation (\ref{eq_topopt_xi_code_compliance}), the \emph{pseudo-energy}, $\xi\bbxhatchi$, can be obtained as
	\begin{equation} \label{eq_topopt_xi_code_complmech}
		\xi\bbxhatchi = -\gamma_1 {\mathbf{\hat{u}}}^{(2)}\bbxhat^{ \text{T}} \left[{\mathbf{B}}\bbxhat^{\text{T}} \overline{\mathbb{C}} {\mathbf{B}}\bbxhat \right] {\mathbf{\hat{u}}}^{(1)}\bbxhat \mdot
	\end{equation}	

\section{MATLAB implementation} \label{sec_matlab_impl}

	The user can run the code from the Matlab prompt with the following Matlab call
	\begin{lstlisting}[language=matlab,numbers=none]
	UNVARTOP_2D_compliance (nelx,nely,nsteps,Vol0,Vol,k,tau)\end{lstlisting}
	where \lstinline{nelx} and \lstinline{nely} stand for the number of quadrilateral elements in the horizontal and vertical directions, respectively.\footnote{The design domains are assumed to be rectangular domains discretized with quadrilateral unit square finite elements.} The following four parameters define the time evolution of the optimization procedure, being \lstinline{nsteps} the number of increments to get from the initial void volume (\lstinline{Vol0}) to the final void volume (\lstinline{Vol}), and parameter \lstinline{k} defines the curvature of the exponential function, in case this type of time-advancing sequence is preferred. For an equally-spaced pseudo-time advance, set \lstinline{k} to 0. The remaining input variable, \lstinline{tau}, rules the minimum filament's width of the optimal design. Other variables related with the topology optimization algorithm and the numerical example (geometry and boundary conditions) are defined inside the function (see Appendix \ref{app_matlab_code}), and can be modified if needed. 
	
	\begin{figure}[pt]
		\centering
		\includegraphics[width=8cm]{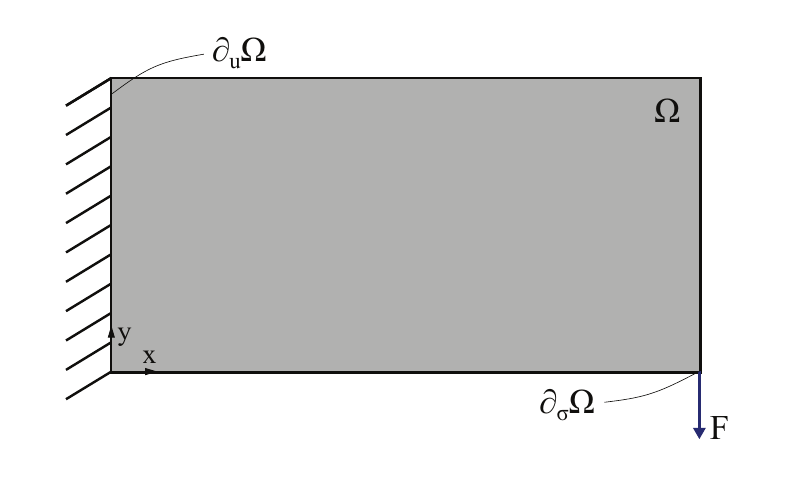}
		\caption{Cantilever beam: topology optimization domain and boundary conditions.}
		\label{fig_cant_domain}
	\end{figure}

	\begin{figure}[pb]
		\includegraphics[width=8cm]{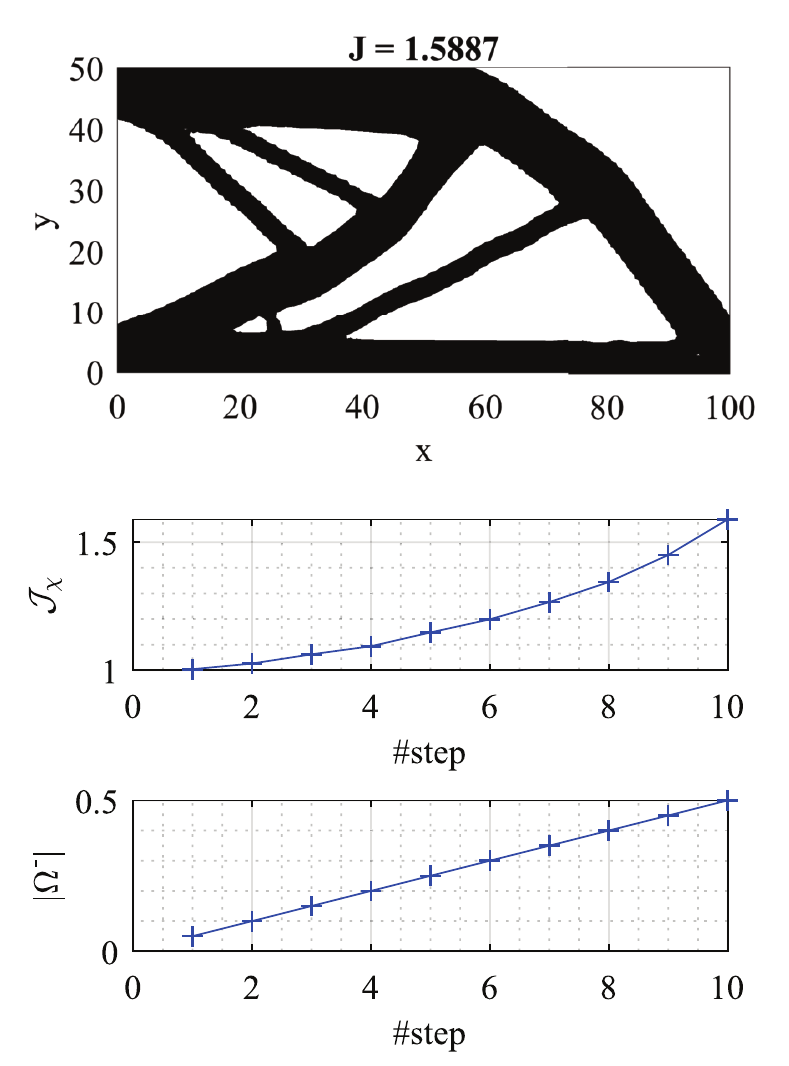}
		\caption{Cantilever beam: topology optimization results.}
		\label{fig_cant_result}
	\end{figure}
	
	For instance, the code can be called with the input line
	\begin{lstlisting}[language=matlab,numbers=none]
	UNVARTOP_2D_compliance (100,50,10,0,0.5,0,0.5)\end{lstlisting}
	for the default example, which corresponds to a cantilever beam with a vertical load applied on the bottom-right corner of $\Omega$, and the displacements are prescribed on the left side of it, as illustrated in Figure \ref{fig_cant_domain}. The algorithm generates two output figures, the first one displays the optimal topology for each iteration, and the second one shows the evolution of the cost function $J_\chi$ and the void volume, $|\Omega^-|$ along the time-steps, as depicted in Figure \ref{fig_cant_result}. At the end, a graphical user interface (GUI) with topology evolution, animated in \href{https://github.com/DanielYago/UNVARTOP/blob/master/Online_Resources/ESM_01.gif}{Online Resource 1}, is shown. 
	
	Relevant details of the Matlab code are explained in the following subsections for the minimum mean compliance problem (section \ref{sec_thry_compliance}), referring to the code in Appendix \ref{app_matlab_code}, along with the required modifications to solve the topology optimization problems defined in sections \ref{sec_thry_multicompliance} and \ref{sec_thry_mechanism}.
		
	\subsection{Parameter definition: lines 2-4} \label{sec_code_parameters}
	
		Table \ref{tab_variables} shows the list of variables and fields required by the program and used along it, excluding the variables already defined in the previous section. These parameters can be grouped in three blocks: all the parameters of the first block are related to the physical problem and the finite element used in the FEM analysis, the next three parameters conform the second block, which define the threshold iterations of the algorithm, and the last one defines a structure of optional parameters to choose which graphics are displayed and which solver is used to solve the \emph{Laplacian regularization}.
			
		\begin{table}[h!]
			\centering
			\caption{List of fields used in the code.}
			\label{tab_variables}
			\begin{tabular}{ ||m{2.1cm}|m{1.5cm}|m{3cm}|| }
				\hline
				Variable & Value & Definition \\
				\hline
				\hline
				\rowcolor{Gray}
				\lstinline|n_dim| & 2 & number of dimensions of the problem \\
				\hline
				\rowcolor{Gray}
				\lstinline|n_unkn| & 2 & number of unknown per node \\
				\hline
				\rowcolor{Gray}
				\lstinline|n_nodes| & 4 & number of nodes per element (e.g. 4 nodes for the quadrilateral element) \\
				\hline
				\rowcolor{Gray}
				\lstinline|n_gauss| & \{1, 4\} & total number of quadrature points of the quadrilateral element \\
				\hline
				\rowcolor{Gray}
				\lstinline|n| & \lstinline|(nelx+1)*(nely+1)| & total number of nodes \\
				\hline
				\rowcolor{Gray}
				\lstinline|h_e| & 1 & element's size \\
				\hline
				\rowcolor{Gray}
				\lstinline|alpha0| & 1e-3 & Prescribed value of $\psi$ for active/passive nodes\\
				\hline
				\rowcolor{LightGray}
				\lstinline|iter_max_step| & 20 & maximum number of in-step iterations \\
				\hline
				\rowcolor{LightGray}
				\lstinline|iter_min_step| & 4 & minimum number if in-step iterations \\
				\hline
				\rowcolor{LightGray}
				\lstinline|iter_max| & 500 & maximum number of iterations \\
				\hline
				\rowcolor{LighterGray}
				\lstinline|opt.Plot_top_iso| & \{true, false\} & Boolean variable to plot the topology along iterations \\
				\hline
				\rowcolor{LighterGray}
				\lstinline|opt.Plot_vol_step| & \{true, false\} & Boolean variable to plot the evolution of the volume along iterations \\
				\hline
				\rowcolor{LighterGray}
				\lstinline|opt.EdgeColor| & \{'none', RGBcolor\} & RGB color of the sides of the quadrilateral elements \\
				\hline
				\rowcolor{LighterGray}
				\lstinline|opt.Solver_Lap| & \{'direct', 'iterative'\} & method to solve the Laplacian regularization \\
				\hline												
			\end{tabular}
		\end{table}	
		
	\subsection{Geometry definition: lines 5-9}
	
\begin{figure}[pb]
			\includegraphics[width=8cm]{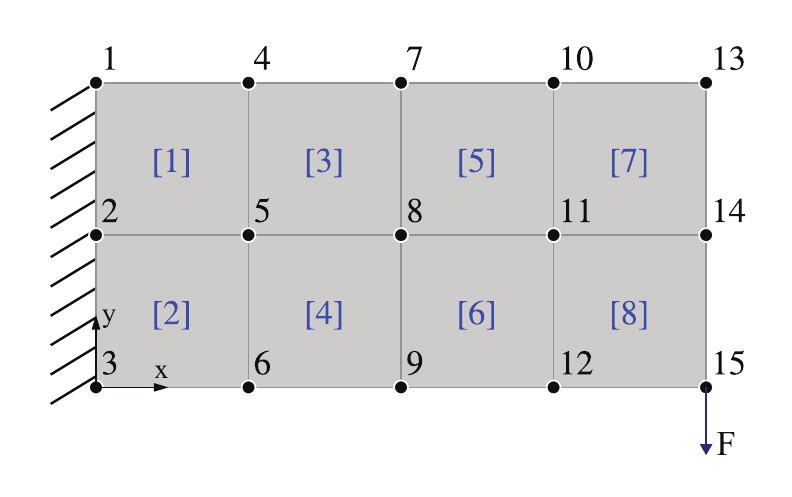}
			\caption{Cantilever beam: mesh discretization.}
			\label{fig_cant_mesh}
		\end{figure}

		The design domain, as aforementioned, is assumed to be rectangular and discretized with square elements. The FE mesh is defined via the coordinates and connectivities arrays, named \lstinline|coord| and \lstinline|connect| in the code. A coarse example mesh of the default example, see Figure \ref{fig_cant_domain}, is illustrated in Figure \ref{fig_cant_mesh}, consisting of 15 nodes and 8 elements, numbered in column-wise (top to bottom) from left to right. The position of each node is defined respect to Cartesian coordinate system with origin at the left-bottom corner.
		
		The \lstinline|coord| matrix is generated using Matlab's \lstinline|meshgrid| function and then, the obtained \lstinline|X,Y| matrices are reshaped into the coordinates matrix, which dimensions are \lstinline|[n x n_dim]|, i.e.
		\begin{equation}
			\text{coord} = \left[  
			\begin{array}{ccccccccc}
			0 & 0 & 0 & 1 & 1 & \dots & 4 & 4 & 4 \\
			2 & 1 & 0 & 2 & 1 & \dots & 2 & 1 & 0 \\
			\end{array}
			\right]^{\text{T}} \mdot
		\end{equation}
		
		The connectivity matrix, \lstinline|connect|, is constructed following the same procedure for computing the degree of freedom connectivity matrix, \lstinline|edofMat|, described by \citet{Andreassen2010}. First, matrix \lstinline|nodenrs| is created with node IDs in a \lstinline|(nely+1)x(nelx+1)| matrix in line 7, mimicking the numbering in Figure \ref{fig_cant_mesh}. Next, the left-bottom node ID of each element is stored in \lstinline|nodeVec| vector, by using matrix \lstinline|nodenrs|. For the given example, this variables are defined as follows:
		\begin{equation}
			\begin{split}
				&\text{nodenrs} = \left[  
				\begin{array}{ccccc}
				1 & 4 & 7 & 10 & 13 \\
				2 & 5 & 8 & 11 & 14 \\
				3 & 6 & 9 & 12 & 15 \\
				\end{array}
				\right] \; \rightarrow \\
				&\hspace{1.5cm}\rightarrow \text{nodeVec} = [2,3,5,6,8,9,11,12]^{\text{T}} \mdot
			\end{split}
		\end{equation}		
		Finally, thanks to the repetitive structure of the grid, the connectivity table, \lstinline|connect|, can be constructed using only \lstinline|nodeVec| and numbering within an element in anticlockwise order starting from the left-bottom node, which reads as
		\begin{equation}
			\text{connect} = \left[  
			\begin{array}{cccc}
				2 & 5 & 4 & 1 \\
				3 & 6 & 5 & 2 \\
				\vdots & \vdots & \vdots & \vdots \\
				11 & 14 & 13 & 10 \\
				12 & 15 & 14 & 11 \\
			\end{array}
			\right] \mdot
		\end{equation}
		
	\subsection{Load and boundary definition: lines 10-17}
	
		Lines 11-17 define the boundary conditions for the displacement and force field. First, the force vector, \lstinline|F|, and the displacement vector, \lstinline|U|, are initialized in lines 11 and 12, respectively. Next, line 13 assigns the imposed force to the force vector, which corresponds to a downwards force applied at the bottom right corner, as illustrated in Figure \ref{fig_cant_domain}, with a small value to limit the maximum displacement of the structure. The next line defines the prescribed degrees of freedom, and stores them in \lstinline|fixed_dofs|.
		
		Parameters \lstinline|active_node| and \lstinline|passive_node| of line 15 are used to force some nodes to be included in the stiff ($\Omega^+$) and soft ($\Omega^-$) material domains, respectively. It is done via the modification of the \emph{discrimination function}, \lstinline|psi|, as in line 54 for the initialization of the \emph{discrimination function} or in the \emph{bisection algorithm} (line 149), by imposing the value \lstinline|alpha0| or \lstinline|-alpha0|.
		
		Finally, the list with free degrees of freedom is generated and stored in \lstinline|free_dofs| (line 16), and the displacement of \lstinline|fixed_dofs| are prescribed to the corresponding value, e.g. 0.
	
	\subsection{Material definition: lines 18-19}
	
		The material used for the analysis is defined in terms of the Young's modulus \lstinline|E0|, of the stiff phase (material domain) and the Poisson's ratio \lstinline|nu|, $\nu$ (see section \ref{sec_thry_state_eq}). In addition, and as a specific parameter of the algorithm, the coefficient \lstinline|m| is defined and prescribed to \lstinline|m=5| for the minimum mean compliance problem. This coefficient in conjunction with the \emph{contrast factor}, \lstinline|alpha|, is used to compute the corresponding \emph{relaxation factor}, \lstinline|beta|. Notice that a noticeably small \emph{contrast factor} can be imposed for compliance problem. 
	
	\subsection{Animation preparation: lines 20-23}
	
		Lines 21-23 initialize the vectors \lstinline|psi_vec|, \lstinline|chi_vec| and \lstinline|U_vec| to 0, which correspond respectively to the \emph{discrimination function}, the \emph{characteristic function} and the displacement vector. This vectors are used to store the corresponding variables at the convergence of each time-step (line 103), and are later called by the \lstinline|Topology_ evolution| GUI.
	
	\subsection{Finite element analysis preprocessing: lines 24-40}
		
		As already mentioned, the regularity in the mesh is highly exploited when computing the global stiffness matrix, \lstinline|K|, to reduce the computational time inside the optimization loop. For that reason, only two element stiffness matrices are required, one for the mixed elements\footnote{The elements bisected by the zero-level of the \emph{discrimination function} are sub-integrated with a single quadrature point according a three-field ($\pmb{\varepsilon}$-$\pmb{\sigma}$-$\hat{\mathbf{u}}$) mixed element. Further details can be found in \cite{Oliver2019}.} and another for the other elements. The first one, is computed with a central quadrature point \lstinline|posgp1|, while the second one requires at least 4 quadrature points to be correctly integrated, \lstinline|posgp4|. The weights of each point are stored in \lstinline|W1| and \lstinline|W4|, respectively. This information is computed by evoking \lstinline|gauss_points| function (lines 111-114) with the total number of point inside the quadrilateral element, as will be later explained.
		
		Next, the nominal constitutive tensor \lstinline|DE| for \lstinline|E0| and \lstinline|nu|, assuming plane-stress (equation (\ref{eq_thry_Pstress})), is computed in line 27, by calling \lstinline|D_matrix_stress| function. The element stiffness matrix, evaluated as
		\begin{equation}
			K_e = \int_{\Omega_e} \mathbf{B}^{\text{T}} \ \overline{\mathbb{C}} \ \mathbf{B} \domega = \sum_{i=1}^{n_{gauss}} w_i |J_i| \mathbf{B}_i^{\text{T}} \ \overline{\mathbb{C}}_i \ \mathbf{B}_i \mcolon
		\end{equation}
		is computed in lines 28-34 for solid and void elements, \lstinline|KE|. The equivalent nominal matrix, for bisected elements, \lstinline|KE_cut| is computed in lines 35-37. The strain-displacement matrix $\mathbf{B}$, defined in lines 119-124 (\lstinline|B_matrix|), is evaluated in each gauss point along with the corresponding determinant of the Jacobian. Moreover, the product $\mathbf{B}^{\text{T}} \, \overline{\mathbb{C}} \, \mathbf{B}$ for the i-th gauss point is stored in \lstinline|KE_i| and \lstinline|K_cut|, respectively.
		
		Finally, the connectivity table of DOFs, \lstinline|edofMat|, is generated in line 38 using built-in \lstinline|kron| and \lstinline|repmat| functions. Each row represents the degrees of freedom of a different element, e.g.
		\begin{equation}
			\text{edofMat} = \left[  
			\begin{array}{cccccccc}
			3 & 4 & 9 & 10 & 7 & 8 & 1 & 2 \\
			5 & 6 & 11 & 12 & 9 & 10 & 3 & 4 \\
			\vdots & \vdots & \vdots & \vdots & \vdots & \vdots & \vdots & \vdots \\
			21 & 22 & 27 & 28 & 25 & 26 & 19 & 20 \\
			23 & 24 & 29 & 30 & 27 & 28 & 21 & 22 \\
			\end{array}
			\right] \mdot
		\end{equation}
		This matrix is now used to compute the indices \lstinline|iK| and \lstinline|jK| used to generate the global stiffness matrix as a sparse matrix from the triplets \lstinline|iK|, \lstinline|jK| and \lstinline|sK|, as will be explained later. 
		
		\subsubsection{Gauss points, Shape function and Cartesian derivatives: lines 110-125}
		
			The bilinear quadrilateral element is used in the FE analysis, which consists of four nodes. Its numerical implementation can be found in the literature \cite{Zienkiewicz2013,Rao2004}.
			
			This element is correctly integrated when 4 quadrature points are employed. The position and weights are computed in \lstinline|gauss_points| function (lines 111-114), where the gauss quadrature points in one direction (parent dimension) are extended to two dimensions, using Matlab's \lstinline|meshgrid| function. \lstinline|posgp| defines the position $[\xi,\eta]$ in the parent square element, where each column represent a different point. The weight values are stored in \lstinline|W| as a row vector. 
			
			The shape matrix, $\mathbf{N}$, (size \lstinline|n_nodes x n_gauss|) is computed in lines 116-117 inside \lstinline|N_matrix| function, as explained in \cite{Zienkiewicz2013,Rao2004}.
			
			Last, the shape derivatives (size \lstinline|n_dim x n_nodes|), the Jacobian matrix $J$ (size \lstinline|n_dim x n_dim|) and the Cartesian derivatives (size \lstinline|n_dim x n_nodes|) are obtained in \lstinline|B_matrix| for a given \lstinline|gauss_points| (lines 119-124), assuming a square unit element. Finally, the strain-displa\-cement matrix $\mathbf{B}$ for the case of interest is computed (size \lstinline|3 x n_nodes*n_unkn|).
			
		\subsubsection{Element Stiffness matrix: lines 125-135} \label{sec_elem_stiff_mat_code}
			
			The constitutive tensor of each element depends on the material properties, which are common to all the elements, and the \emph{characteristic function}. Due to this regularity, the nominal constitutive tensor, $\overline{\mathbb{C}}$, is computed only once for the stiff material properties, in lines 126 and 127, and the corresponding stiffness matrix is later multiplied by the term $\chi_\beta ^m$, which depends on each element. The \emph{relaxed characteristic function} is calculated in lines 129-131, inside \lstinline|interp_property| function, as follows
			\begin{equation*}
				coeff = \left\{
				\begin{split}
					&\left(\chi + (1-\chi) \beta\right)^m \, && \text{for stiffness} \\
					m&\left(\chi + (1-\chi) \beta\right)^{m-1} (1-\beta) \, && \text{for sensitivity} 
				\end{split}
				\right.
			\end{equation*}
			with $\chi\in[0,1]$.
			
			The global stiffness matrix is assembled at each iteration inside the optimization loop using Matlab's \lstinline|sparse| function to addition the components with same i-th (\lstinline|iK|) and j-th (\lstinline|jK|) degree of freedom, calling \lstinline|assembly_stiff_mat|. Its definition is written in lines 133-135, where the third component (\lstinline|sK|) for the \lstinline|sparse| function is computed. Each column of the \lstinline|sK| matrix corresponds to the stiffness matrix of element \lstinline|e|. It is worth emphasizing that the bisected elements must be multiplied by \lstinline|KE_cut|.
	
	\subsection{Laplacian regularization preparation: lines 41-52} \label{sec_code_lap_reg}
	
		Mimicking the preprocessing procedure of the global stiffness matrix (see section \ref{sec_elem_stiff_mat_code}), the lhs matrix of equation \ref{eq_regularized_laplacean_smoothing_psi} can be computed just once (lines 42-48), since it does not depend on the topology but on the mesh, which is regular. Thus, the terms $\nabla\mathbf{N}^{\text{T}} \nabla\mathbf{N}$ and $\mathbf{N}^{\text{T}} \mathbf{N}$, which correspond to \lstinline|KE_Lap| and \lstinline|ME_Lap| defined in lines 42 and 43, are analytically computed and defined as
		{\allowdisplaybreaks
		\begin{align*}
				& KE_{Lap} = \int_{\Omega} \mathbf{B}^{\text{T}} \ \mathbf{B} \domega \rightarrow KE_{Lap} = \dfrac{1}{6} \left[  
				\begin{array}{rrrr}
				4 & -1 & -2 & -1 \\
				-1 & 4 & -1 & -2 \\
				-2 & -1 & 4 & -1 \\
				-1 & -2 & -1 & 4 \\
				\end{array}
				\right]\\
				& ME_{Lap} = \int_{\Omega} \mathbf{N}^{\text{T}} \ \mathbf{N} \domega \rightarrow ME_{Lap} = \dfrac{1}{36} \left[  
				\begin{array}{rrrr}
				4 & {\color{white}-}2 & {\color{white}-}1 & {\color{white}-}2 \\
				2 & 4 & 2 & 1 \\
				1 & 2 & 4 & 2 \\
				2 & 1 & 2 & 4 \\
				\end{array}
				\right] \mdot
		\end{align*}}
		Next, combining both matrices and the \emph{regularization parameter} $\tau$, the lhs matrix is generated and saved in \lstinline|KE_Lap| (line 44). Lines 45 to 48 define the triplets \lstinline|i_KF|, \lstinline|j_KF| and \lstinline|s_KF|, which are then used to obtain the sparse matrix \lstinline|K_Lap| in line 48.
		
		Depending on \lstinline|opt.solver_Lap|, the Laplacian regularization will be solved using a direct or an iterative method. This procedure can be sped up by computing the Cholesky factorization of the lhs (\lstinline|chol(K_Lap,'lower')|) if the direct method is chosen, or computing the incomplete Cholesky factorization (\lstinline|ichol(K_Lap,opts)|, with \lstinline|opts = struct('type','ict','droptol',1e-3,'diagcomp',0.1)|) in case an iterative algorithm is desired. It will be later used as a preconditioner. 
		
		The rhs must be computed at each iteration, since it depends on the \emph{discrimination function}, \lstinline|psi|, as detailed in section \ref{sec_lapl_reg_code}. Nevertheless, the resolution procedure of equation \ref{eq_regularized_laplacean_smoothing_psi} can be prepared by computing both the shape matrix, \lstinline|N_T|, of size \lstinline|n_nodes x n_gauss| \footnote{The matrix is transposed with respect to the common one.}, and the indexes of the element nodes \lstinline|i_xi| (reshaping the connectivity matrix into a column vector). The assembly is carried out in lines 83 and 85 evoking \lstinline|accumarray| function.		
		
	\subsection{Main program: lines 53-107}
	
		The main optimization procedure starts by initializing the topology via the \emph{discrimination function} to \lstinline|alpha0|, constant to all the nodes, except for those listed in \lstinline|passive_node|. Next, the \emph{characteristic function} is obtained via \lstinline|compute_volume| function. Line 57 is used to initialize several vectors, which will accumulate the convergence variables (cost function, volume and lambda), and other essential variables. The initial topology is displayed in the next line by means of \lstinline|plot_isosurface|.
		
		The optimization starts in line 59, where the loop over time-steps is defined. As explained in section \ref{sec_thry_algorithm}, the \emph{reference pseudo-time} is iteratively increased following a linear or exponential expression, which definition is written through lines 186-188, and for each time-step the optimization loop is repeated until convergence is achieved. The optimization loop (lines 65-99) consists of five parts: finite element analysis, sensitivity computation, Laplacian regularization, topology update and convergence check.
		
		Finally, at each iteration, the topology is plotted (line 92),
		the intermediate results are printed in display (line 96) and the iteration counters are increased (line 97).
		
		\subsubsection{Finite element code: lines 66-69} \label{sec_code_FEM}
		
			The global stiffness matrix, \lstinline|K|, is assembled inside \lstinline|assembly _stiff_mat| function using the \lstinline|sparse| function, where \lstinline|sK| is computed considering the corresponding \emph{relaxed characteristic function} for each element. Next, in line 68, the equilibrium equation (\ref{eq_equilibrium}) is solved using a direct method. The displacements are stored in \lstinline|U|. Next, \hypertarget{sentcodecostfunction}{the cost function}, \lstinline|J|, normalized with the one of the first iteration (\lstinline|J_ref|), can be obtained at the current topology layout.
		
		\subsubsection{Sensitivity computation: lines 70-79} \label{sec_code_sensitivity}
		
			According to equation \ref{eq_optimal_condition}, the \emph{energy density} is defined as the partial derivative of the cost objective's kernel multiplied by the exchange function, $\Delta\chi$. The \emph{energy density} is computed in two parts, in the first one (lines 72 to 75) the sensitivity of non-bisected elements is obtained for the 4 quadrature points, while the sensitivity for the mixed elements is calculated in the second part (lines 76 to 78). 
			
			The element sensitivity, as detailed in section \ref{sec_thry_compliance} for the minimum compliance problem, is computed as $m \chi_\beta^{m-1} \, \mathbf{u}_e^{\text{T}} K_{e,i} \mathbf{u}_e \, (1-\beta)$ for the $e$-th element and the $i$-th gauss point (see equation (\ref{eq_topopt_xi_code_compliance})). However, for the bisected elements, the element stiffness matrix $K_{e,i}$ is replaced by \lstinline|K_cut|, and the resultant value is copied to the four gauss points.
				
			At the first iteration, the parameters 	\lstinline|xi_shift| and \lstinline|xi_norm| are defined as
			\begin{lstlisting}[numbers=none]
	xi_shift = min(0,min(Energy(:))) 
	xi_norm = max([range(Energy(:));Energy(:)])\end{lstlisting}
			and will be used to obtain the \emph{modified energy density}, $\hat{\xi}\bbx$, described in equation (\ref{eq_thry_shift_norm}).
				
		\subsubsection{Laplacian regularization: lines 80-86} \label{sec_lapl_reg_code}
		
		As aforementioned, instead of applying the \emph{Laplacian regularization} (\ref{eq_regularized_laplacean_smoothing_psi}) to the resultant \emph{discrimination function}, at each iteration of the \emph{bisection algorithm} and since it does not affect constant fields such as $\lambda$, the \emph{Laplacian regularization} is only implemented for $\hat{\xi}$. The corresponding system is defined as
		\begin{equation} \label{eq_regularized_laplacean_smoothing}
			\left\{
			\begin{split}
				&\xi_{\tau}-(\tau h_e)^2\Delta_{\mathbf x}\xi_{\tau}=\hat{\xi}& &\quad in \; \Omega\\
				&\nabla_{\mathbf x}\xi_{\tau}\cdot\mathbf{n}={0}& &\quad on \; \partial\Omega
			\end{split}
			\right. \mcolon
		\end{equation}
		where $\hat{\xi}$ and $\xi_{\tau}$ stand for the \emph{modified unfiltered energy density} and the \emph{smooth energy density}, respectively. As commented in section \ref{sec_thry_formulation}, the lhs has been precomputed (see section \ref{sec_code_lap_reg}) and the rhs is now computed based on the \emph{modified energy density} (field on gauss points). FE discretization of the rhs leads to 
		\begin{equation}
			rhs = \int_{\Omega} {\mathbf N}^{\text{T}} \hat{\xi} \bbx \domega \mcolon
		\end{equation}
		which can be rewritten in matrix form, defined in line 81, as
		\begin{lstlisting}[firstnumber=81,tabsize=1,gobble=3]
			xi_int = N_T*(Energy-xi_shift*chi)/xi_norm;\end{lstlisting}
		The nodal contribution of this integral is later constructed by means of the built-in \lstinline|accumarray|	Matlab function.			
		
		The system of linear equations (\ref{eq_regularized_laplacean_smoothing}), as mentioned, can be solved using the Cholesky factorization and a direct solver (line 83) or an iterative solver (e.g. \lstinline|minres| solver) applying the incomplete Cholesky factorization as the preconditioner of the system, as described in section \ref{sec_code_lap_reg}.
		
		It is worth to mention that for low number of elements, as it is the case of this paper since it is for academic purposes, the \emph{Laplacian regularization} may generate boundary waves for thin filaments, as displayed in some figures. This undesirable effect should vanish if finer meshes are used.
				
		\subsubsection{Update of $\chi$ and $\psi$: line 88} \label{sec_code_update_top}
		
			The topology layout, satisfying the constraint equation, is obtained by means of a \emph{bisection algorithm} (solution of equation (\ref{eq_solution_psi_xi_2})) called in line 88. The \lstinline|find_volume| functions computes the Lagrange multiplier $\lambda$ (\lstinline|lambda|), the new \emph{discrimination function} $\psi$ (\lstinline|psi|) and the corresponding \emph{characteristic function} $\chi$ (\lstinline|chi|).
		
			\paragraph{Bisection algorithm: lines 137-152} \label{sec_code_bisection}
			
				The \emph{bisection algorithm} consists of a search for a suitable bracket, and the subsequent root finding. The left and right extremes of the interval are easily defined by the minimum and maximum value of the energy density field, and stored as \lstinline|l1| and \lstinline|l2|, respectively. The corresponding constraint values are saved as \lstinline|c1| and \lstinline|c2|. Lines 139-141 tests the last $\lambda$ as a trial extreme of the interval, by means of \lstinline|compute_volume_lambda|, to reduce the number of iterations. The bisection loop is written in lines 142 to 146, where the root of the constraint equation is estimated as the midpoint of the bracketing interval (line 143). 
				
				At each iteration of the bisection, given a \emph{density function} \lstinline|xi| and a trial \lstinline|lambda|, the \emph{discrimination function} is obtained at line 149. The active and passive nodes are considered by modifying the \lstinline|psi| function, as aforementioned. The void volume ratio \lstinline|vol| and the \emph{characteristic function} \lstinline|chi| are obtained from \lstinline|compute_volume| in line 150. Next, the constraint equation is evaluated in line 151, and the extremes of the interval are updated accordingly. This procedure is repeated until the void volume is within $10^{-4}$ of the reference time.
					
			\paragraph{Volume computation: lines 153-161}
			
				The computation of the volume is done by means of an integration with 36 quadrature points. This methodology differs for simplicity of the implementation from the one used in \citet{Oliver2019}, where a modified marching squares was employed. 
				
				The position and weights of the 36 quadrature points are assigned as defined in \cite{Rao2004}. First, line 158 determines which elements are bisected by the internal boundary through the nodal value of the \emph{discrimination function}. In case they all have the same sign, the boundary will not cross throughout the element. 
				Then, the element nodal $\psi$, \lstinline|psi_n|, is evaluated in the quadrature points for the bisected elements and saved as  \lstinline|psi_x|. The \emph{characteristic function} is obtained as the dot product of \lstinline|W| and \lstinline|phi_x>0|. Finally, the void volume ratio is computed in line 161. 
		
		\subsubsection{Convergence check: lines 90 and 94} \label{sec_code_convergence}
		
			Lines 90 and 94 compute the convergence tolerances of the algorithm, along with the constraint tolerance \lstinline|Tol_constr|. The convergence is checked inside the while condition at line 65, and it only converges when the number of in-step iterations (\lstinline|iter_step|) is in between \lstinline|iter_min_step| and \lstinline|iter_max_step|, and the three following conditions are satisfied: the L2-norm of the \emph{characteristic function} is less than 0.1, the relative difference of the Lagrange multiplier with respect the previous one is less than 0.1, and the volume magnitude is within $10^{-4}$ of the desired \emph{pseudo-time}, \lstinline|t_ref|. 
			
			The optimization terminates if the maximum number of iterations, \lstinline|iter_max|, or the maximum number of in-step iterations are achieved, showing a warning message in command window. 
		
	\subsection{Iso-surface plot: lines 92 (162-172)}
		
		The \lstinline|plot_isosurface| function shows the optimal topology via the \emph{discrimination function} \lstinline|psi| in a black-and-white design, as seen in Figure \ref{fig_cant_result} (top view). The behavior of this function depends on the iteration, i.e. the first time it is called, a figure is generated and its handle saved as \lstinline|fig_handle|. In addition, the topology is represented using built-in \lstinline|patch| function, the handle of which is stored as \lstinline|obj_handle|, by means of the coordinates matrix, connectivity matrix and the nodal \emph{discrimination function}. However, only \lstinline|psi| field is updated in the other iterations using \lstinline|set(obj_handle,'FaceVertexCData',psi);|. 
		
	\subsection{Cost function and volume vs. step plot: line 102 (173-184)} \label{sec_code_plot_cost}
	
		The definition of \lstinline|plot_volume_iter| function is similar to that of \lstinline|plot_isosurface| function. At iteration 1, it creates the figure, and two axes using \lstinline|subplot| function. The cost function evolution is illustrated in the top subplot, while the volume evolution is displayed at the bottom. At other iterations, the lines are updated using \lstinline|set| function, with the updated \lstinline|J_vec| and \lstinline|vol_vec| vectors, respectively.
	
	\subsection{Topology evolution GUI: lines 108-109} \label{sec_code_plot_gui}
	
		Once the Topology Optimization problem has been solved, the results can be graphically post-processed by means of a graphical user interface (GUI), where the topology and displacement fields are displayed for the set of time-steps. It is created by the following function call:
\begin{lstlisting}[language=Matlab,firstline=127,lastline=127,numbers=none]
Topology_evolution(coord,connect,[Vol0,vol_vec],psi_vec,chi_vec,U_vec);
\end{lstlisting}
		where \lstinline{vol_vec} corresponds to the set of \emph{pseudo-time} values for which the topology has been optimized. Then, \lstinline{psi_vec} and \lstinline{chi_vec} correspond respectively to the \emph{discrimination function} (nodal scalar field) and the \emph{characteristic function} (element scalar field), each column corresponding to a different time-step. Similarly, \lstinline{U_vec} correspond to the displacement field, where each column and layer of the array represent a different loading condition and a different time-step, respectively.
		
		\begin{figure}[pt]
			\includegraphics[width=8cm]{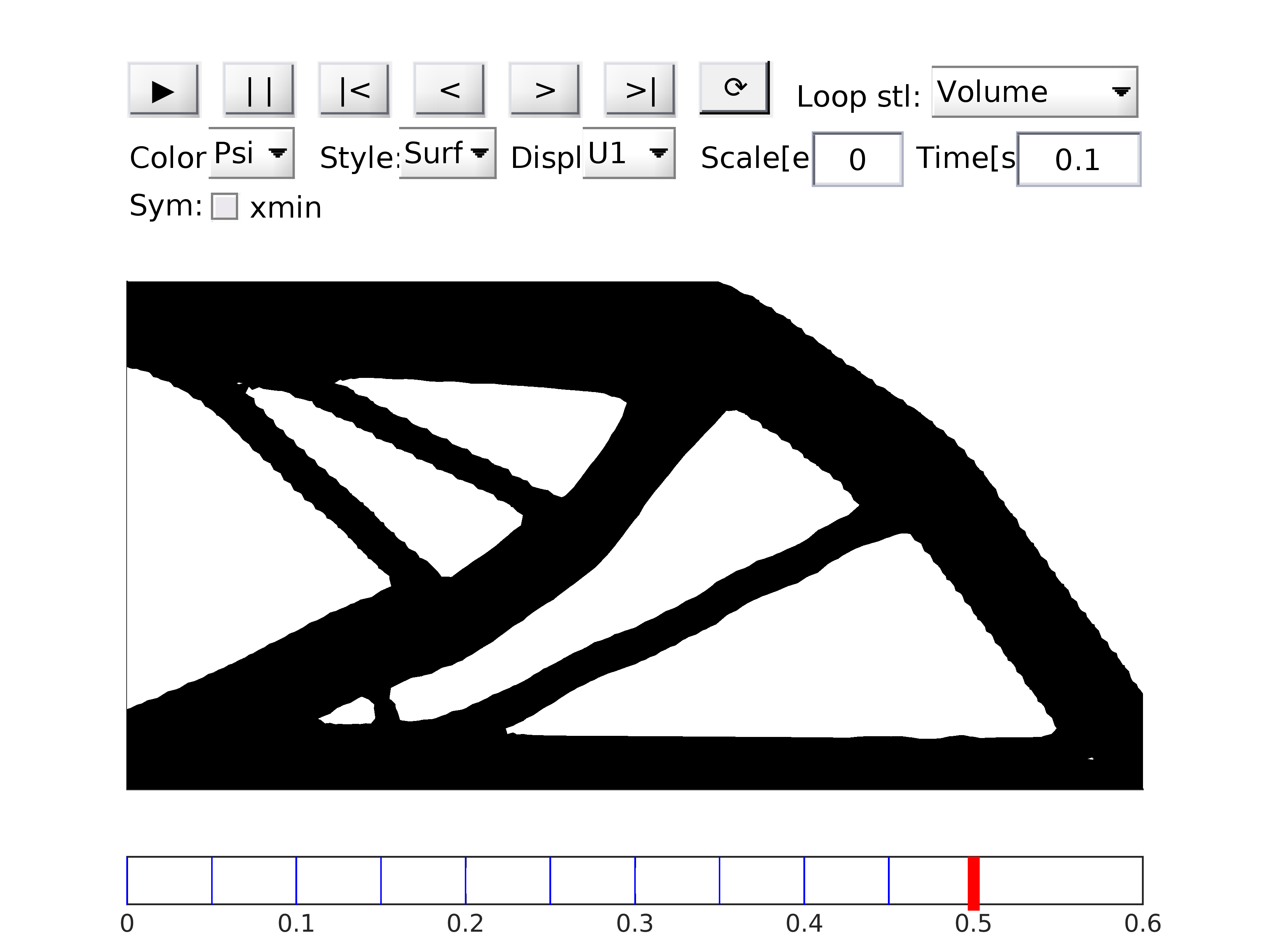}
			\caption{GUI's design.}
			\label{fig_GUI}
		\end{figure}
		
		\begin{figure*}[pb]
			\centering
			\includegraphics[width=17cm]{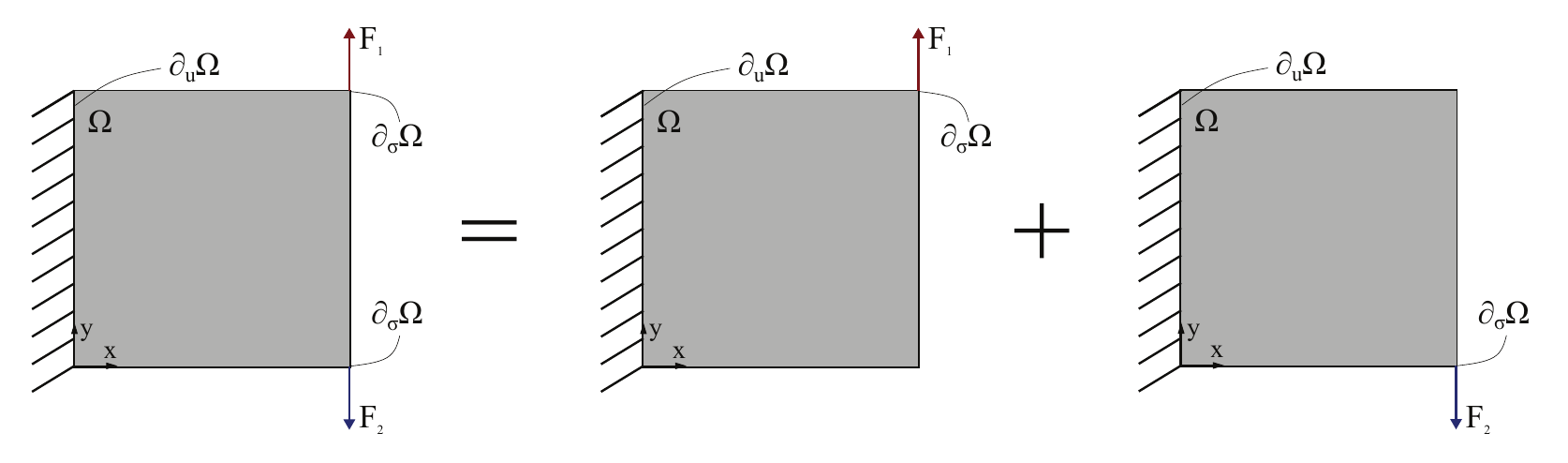}
			\caption{Multi-load beam: topology optimization domain and boundary conditions.}
			\label{fig_multiload_domain}
		\end{figure*}		
		
		The interface allows to select the field to display (psi, chi or the norm of the displacement for any load condition) and the style of the representation (surface only, wireframe only and surface plus wireframe). The user can also choose the scale factor and the displacement field to deform the mesh as it can be observed in Figure \ref{fig_GUI}.
			
		The set of push-buttons on the top-left area controls the animation of the topology along the \emph{pseudo-time}, the time between time-steps can be modified in the $dt$ text edit field. The last button corresponds to a toggle-button, which animates indefinitely the topology until it is clicked. Depending on the chosen loop style option, the topology is animated along the time-steps (Volume) or along the scale factor for a given time-step (Scale linear and Scale sine).
		
		The possibility to mirror/symmetrize the topology is the last relevant feature of this figure. A set of checkboxes allow to symmetrize the mesh and its properties on any of the sides of the domain.
	
	\subsection{Multi-load mean compliance: code modification} \label{sec_code_multi_load}

		According to section \ref{sec_thry_multicompliance}, the program can be easily adapted to optimize multi-load problems, as shown in Figure \ref{fig_multiload_domain}. Then, the cost function as well as the sensitivity are evaluated as weighted averages of each individual optimization problem.
			
		First, the loads and boundary conditions are changed to include the second loading state\footnote{More than one additional loading state can be considered.}, defined in the second column of \lstinline|F|:
\begin{lstlisting}[firstline=22,lastline=23,firstnumber=11,belowskip=0pt]
F = sparse(n_unkn*n,2);
U = zeros(n_unkn*n,2);
\end{lstlisting}
\begin{lstlisting}[firstline=24,lastline=25,numberstyle=\tiny\color{codegray}13-,aboveskip=0pt]
F(n_unkn*find(coord(:,2)==nely & coord(:,1)==nelx),1) = 0.01*nelx;
F(n_unkn*find(coord(:,2)==0 & coord(:,1)==nelx),2) = -0.01*nelx;
\end{lstlisting}
		Furthermore, an additional column is added to \lstinline|U_vec| by replacing line 23 with
\begin{lstlisting}[firstline=35,lastline=35,firstnumber=23]
U_vec = zeros(n_unkn*size(coord,1),size(U,2),nsteps+1);
\end{lstlisting}
		
		\begin{figure}[pb]
			\centering
			\includegraphics[width=8cm]{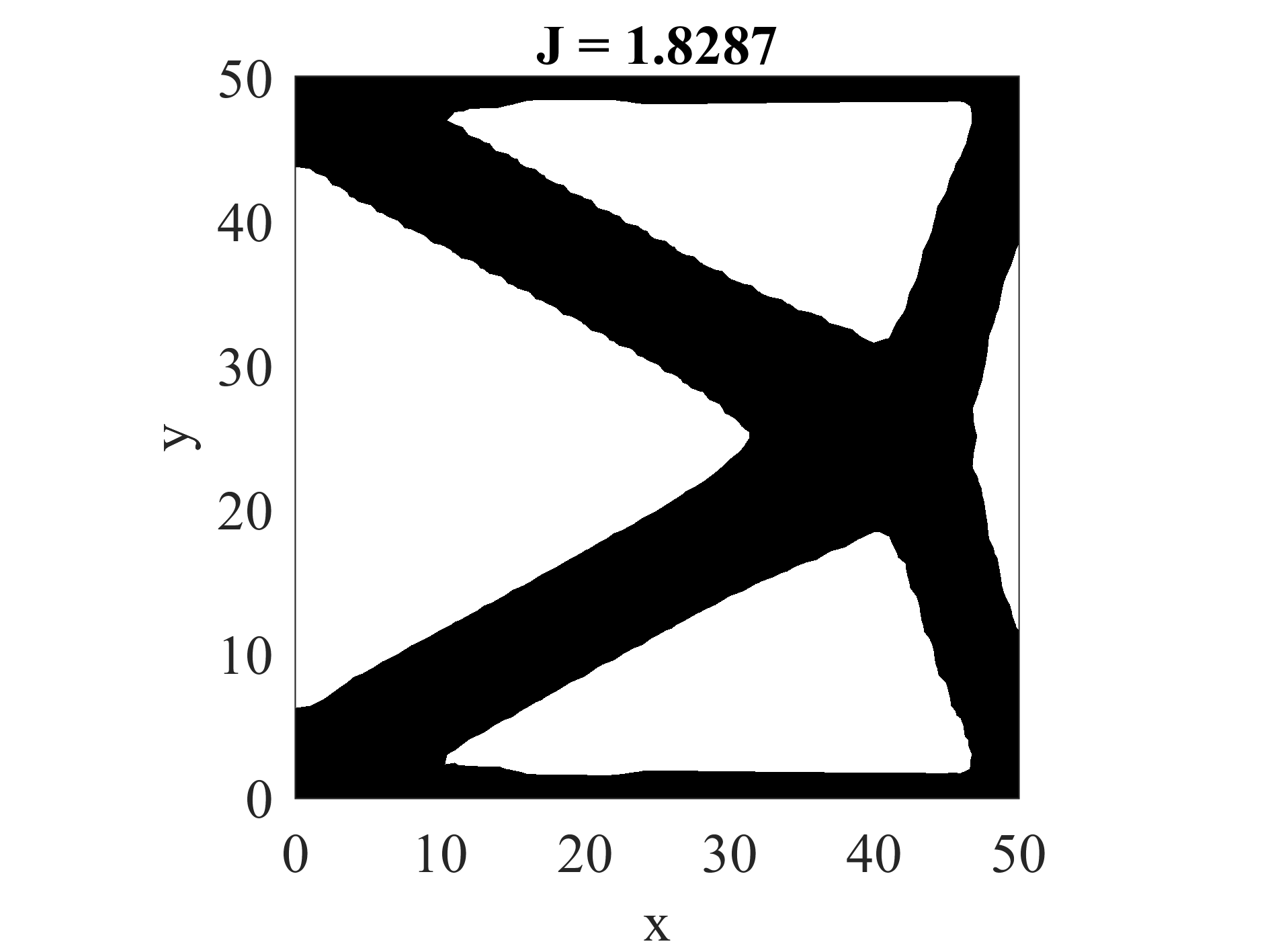}
			\caption{Multi-load beam: optimal topology layout.}
			\label{fig_multiload_topology}
		\end{figure}
		
		Next, the sensitivity computation must be adapted to include multiple loading states, via a \lstinline|for| loop. Then, lines 73-75 are substituted with
\begin{lstlisting}[firstline=85,lastline=86,numberstyle=\tiny\color{codegray}73-,belowskip=0pt]
		for i_load=1:size(F,2)
			u_e = reshape(U(edofMat(id,:)',i_load),n_nodes*n_unkn,[]); w_e = u_e;
\end{lstlisting}
\begin{lstlisting}[firstline=87,lastline=88,firstnumber=74,aboveskip=0pt]
			for i=1:n_gauss; Energy(i,id) = Energy(i,id) + sum(w_e.*(KE_i(:,:,i)*u_e),1); end
		end; Energy(:,id) = int_chi.*Energy(:,id);
\end{lstlisting}
		and equivalently, lines 77 and 78 are replaced by
\begin{lstlisting}[firstline=90,lastline=91,numberstyle=\tiny\color{codegray}77-,belowskip=0pt]
		for i_load=1:size(F,2)
			u_e = reshape(U(edofMat(id,:)',i_load),n_nodes*n_unkn,[]); w_e = u_e;
\end{lstlisting}
\begin{lstlisting}[firstline=92,lastline=93,numberstyle=\tiny\color{codegray}78-,aboveskip=0pt]
			Energy(:,id) = Energy(:,id) + repmat(sum(w_e.*(K_cut*u_e),1),n_gauss,1);
		end; Energy(:,id) = int_chi.*Energy(:,id);
\end{lstlisting}
			
\begin{figure}[pb]
			\centering
			\includegraphics[width=8cm]{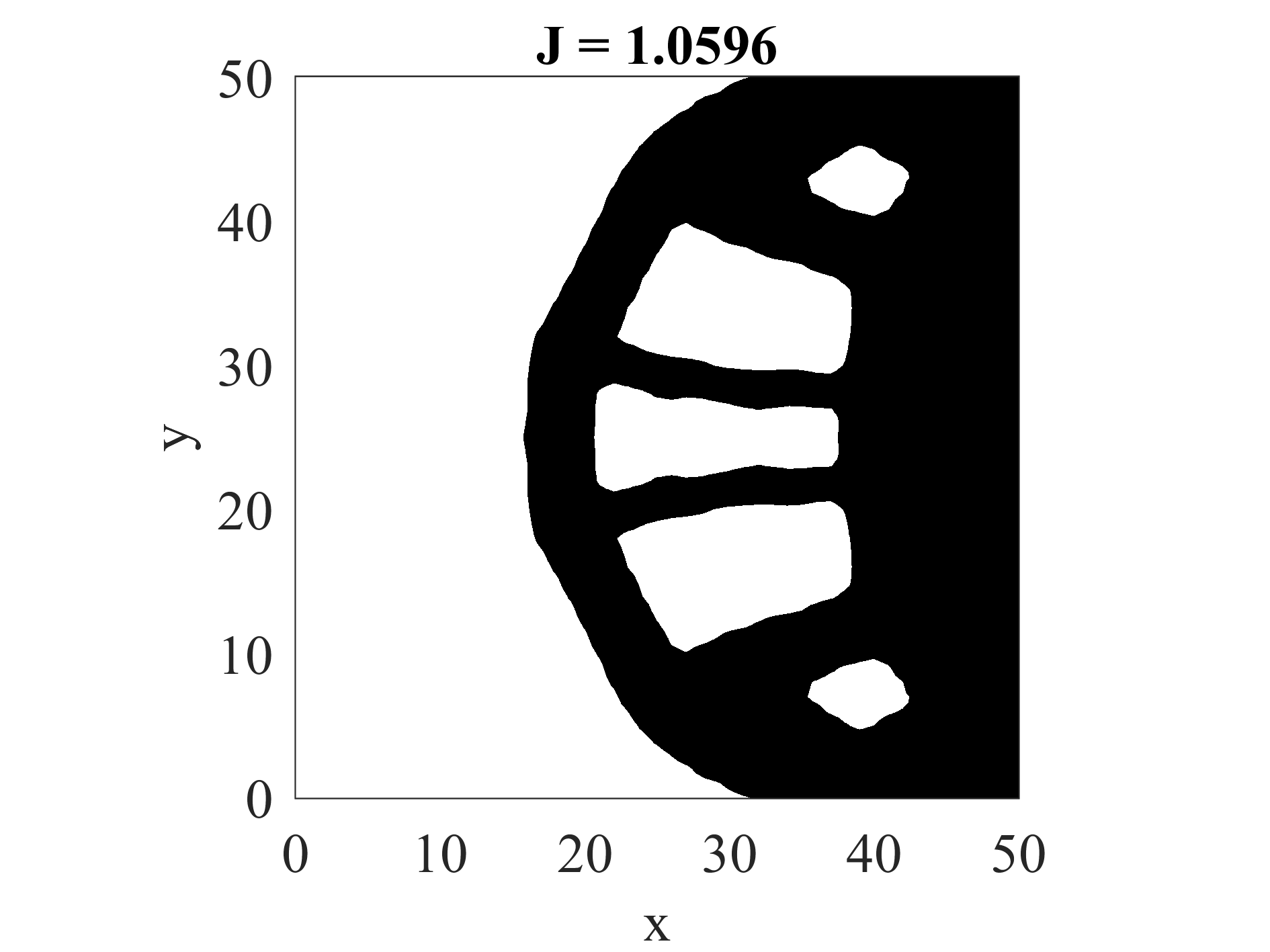}
			\caption{Multi-load beam: optimal topology layout when loads are applied at the same time.}
			\label{fig_singleload_topology}
		\end{figure}

		This example can be simulated by the following line
		\begin{lstlisting}[numbers=none]
			UNVARTOP_2D_multiload (50,50,11,0,0.55,0,0.5)\end{lstlisting}
		The resultant optimal topology, at $t_{ref}=0.55$, is displayed in Figure \ref{fig_multiload_topology}, while the topology evolution is shown in \href{https://github.com/DanielYago/UNVARTOP/blob/master/Online_Resources/ESM_02.gif}{Online Resource 2}. It can be observed in Figure \ref{fig_singleload_topology} how much the topology differs from the single loading condition, when the two loads of Figure \ref{fig_multiload_domain} are applied at the same time.
		
	\subsection{Compliant mechanisms: code modification} \label{sec_code_complmechanism}
	
\begin{figure}[pb]
			\centering
			\includegraphics[width=8.2cm]{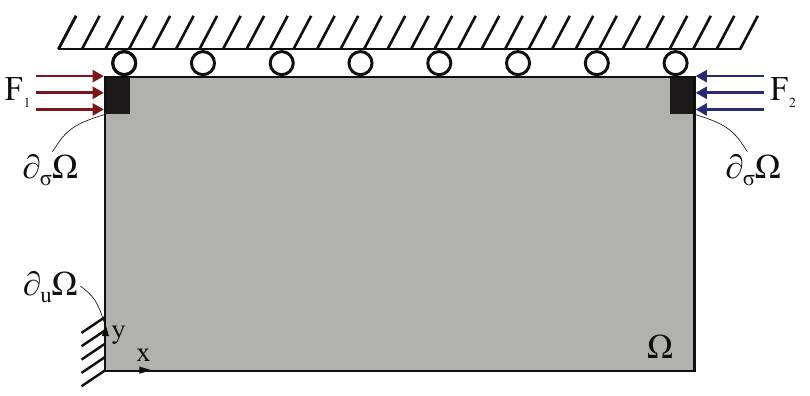}
			\caption{Inverter (compliant mechanism): topology optimization domain and boundary conditions.}
			\label{fig_complmechanism_domain}
		\end{figure}

		Mimicking the previous section, the base code in Appendix \ref{app_matlab_code} also requires some modifications in order to optimize compliant mechanisms, as depicted in Figure \ref{fig_complmechanism_domain}. A second loading state must be solved to compute the \emph{adjoint state} $\mathbf{w}$, which is later used in the sensitivity computation. This second state is loaded with a dummy constant load applied in the \emph{output nodes} in the same direction as the desired displacement. Then, the loads and boundary conditions are modified to
\begin{lstlisting}[firstline=22,lastline=23,firstnumber=11,belowskip=0pt]
F = sparse(n_unkn*n,2);
U = zeros(n_unkn*n,2);
\end{lstlisting}
\begin{lstlisting}[firstline=24,lastline=25,numberstyle=\tiny\color{codegray}13-,aboveskip=0pt,belowskip=0pt]
F(n_unkn*find(coord(:,2)>=0.9*nely & coord(:,1)==0)-1,1) =     0.0001*nelx;
F(n_unkn*find(coord(:,2)>=0.9*nely & coord(:,1)==nelx)-1,2) = -0.0001*nelx;
\end{lstlisting}
\begin{lstlisting}[firstline=26,lastline=27,numberstyle=\tiny\color{codegray}14-,aboveskip=0pt,belowskip=0pt]
fixed_dofs = [reshape(n_unkn*find(coord(:,2)==nely),1,[]),...
	reshape(n_unkn*find(coord(:,1)==0 & coord(:,2)<=0.1*nely)+(-n_unkn+1:0),1,[])];
\end{lstlisting}
\begin{lstlisting}[firstline=28,lastline=29,numberstyle=\tiny\color{codegray}15-,aboveskip=0pt]
active_node = find(coord(:,2)>0.9*nely&(coord(:,1)<0.05*nelx|coord(:,1)>0.95*nelx));
passive_node = [];
\end{lstlisting}
		Notice that the force is applied along a segment, and not only in a single node. Furthermore, only half of the design is computed thanks to the symmetry of the design and some nodes surrounding the \emph{input} and \emph{output ports} are forced to remain as stiff material.
		
		The properties of the material (line 19) should be also changed to \lstinline|m=3| and \lstinline|alpha=1e-2|. This adjustment increases convergence.
		
		To ensure fast convergence, external springs must be included in the \emph{input} and \emph{output ports} at the same degrees of freedom as the applied forces. These degrees are obtained by means of the following lines:
\begin{lstlisting}[firstline=32,lastline=33,numbers=none]
id_in  = find(F(:,1)); id_in  = sub2ind(n_unkn*(nely+1)*(nelx+1)*[1 1],id_in,id_in);
id_out = find(F(:,2)); id_out = sub2ind(n_unkn*(nely+1)*(nelx+1)*[1 1],id_out,id_out);
\end{lstlisting}
		which must be inserted between lines 17 and 18. These two lists are used inside \lstinline|assembly_stiff_mat|, thus its call has to be replaced by
\begin{lstlisting}[firstline=83,lastline=83,firstnumber=67]
		[K] = assmebly_stiff_mat (chi,KE,KE_cut,beta,m,iK,jK,n_unkn,nelx,nely,id_in,id_out);
\end{lstlisting}
		as well as its definition at line 133
\begin{lstlisting}[firstline=151,lastline=151,firstnumber=133]
function [K] = assmebly_stiff_mat (chi,KE,KE_cut,beta,m,iK,jK,n_unkn,nelx,nely,id_in,id_out)
\end{lstlisting}
		
		The external springs, using \lstinline|id_in| and \lstinline|id_out|, are added to the global stiffness matrix after line 135:
\begin{lstlisting}[firstline=154,lastline=155,numbers=none]
K(id_in)  = K(id_in)  + 0.002;
K(id_out) = K(id_out) + 0.002;
\end{lstlisting}
		The prescribed value for the springs must be adjusted for each individual example.
		
		\begin{figure}
			\centering
			\includegraphics[width=8cm]{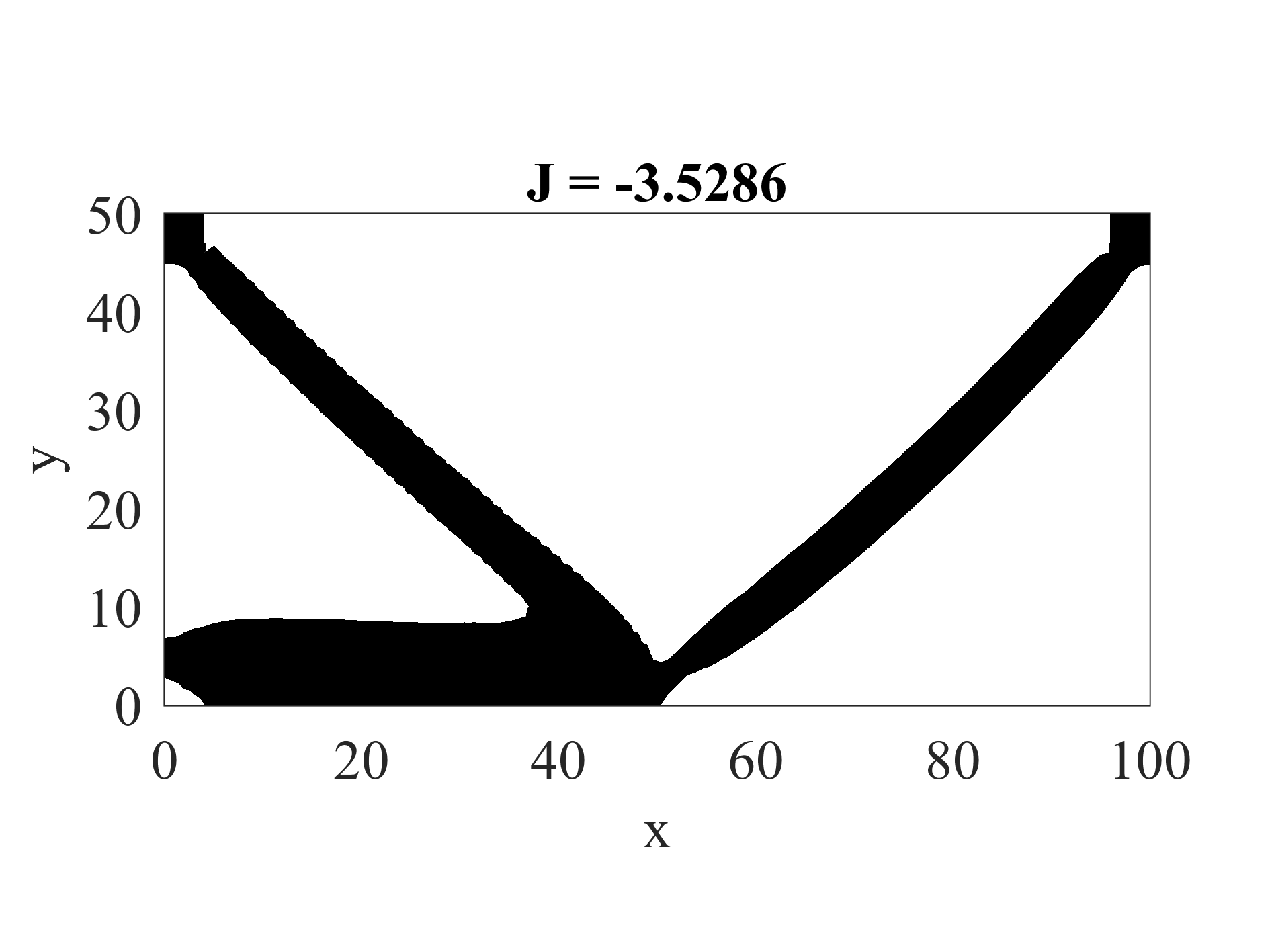}
			\caption{Inverter (compliant mechanism): optimal topology layout.}
			\label{fig_complmechanism_topology}
		\end{figure}
		
		The cost function must be also replaced by the corresponding work at the \emph{output port}, since the cost function is defined as the maximization of the output displacement. It is implemented by the following line:
\begin{lstlisting}[firstline=85,lastline=85,firstnumber=69]
		if iter == 1; U_vec(:,:,1)=U; J_ref = -abs(F(:,2)'*U(:,1)); end; J = F(:,2)'*U(:,1)/J_ref;
\end{lstlisting}
		
		As in section \ref{sec_code_multi_load}, \lstinline|U_vec| must be substituted by
\begin{lstlisting}[firstline=39,lastline=39,firstnumber=23]
U_vec = zeros(n_unkn*size(coord,1),2,nsteps+1);
\end{lstlisting}
		
		Finally, the displacements of the \emph{adjoint system}, used in the calculation of the sensitivity, must be replaced by the corresponding displacements of the second system. Thus, these lines are now defined as
\begin{lstlisting}[firstline=89,lastline=90,numberstyle=\tiny\color{codegray}73-]
		u_e = reshape(U(edofMat(id,:)',1),n_nodes*n_unkn,[]);
		w_e = -reshape(U(edofMat(id,:)',2),n_nodes*n_unkn,[]);
\end{lstlisting}
		and
\begin{lstlisting}[firstline=94,lastline=95,numberstyle=\tiny\color{codegray}77-]
		u_e = reshape(U(edofMat(id,:)',1),n_nodes*n_unkn,[]);
		w_e = -reshape(U(edofMat(id,:)',2),n_nodes*n_unkn,[]);
\end{lstlisting}
				
		The optimal topology, for the given boundary conditions, illustrated in Figure \ref{fig_complmechanism_topology} can be performed with
		\begin{lstlisting}[numbers=none]
			UNVARTOP_2D_complmechanism (100,50,10,0,0.8,-2,0.5)\end{lstlisting}
		The resultant compliant mechanism is animated in \href{https://github.com/DanielYago/UNVARTOP/blob/master/Online_Resources/ESM_03.gif}{Online Resource 3}.

\section{Numerical examples} \label{sec_num_examples}
	
	The following numerical examples exhibit the potential of the unsmooth variational topology optimization technique in 2D problems. Unless otherwise stated, the parameters and material properties are left as the default examples, for each of the three optimization problems described in this work. The design domain, the function call and the boundary conditions for each example are defined in Table \ref{tab_example}.
	
	\begin{table*}
		\topcaption{List of examples.}\label{tab_example}
		\begin{supertabular}{ ||c m{4cm} m{8cm}|| }
			\hline
			\hline
			Domain & Matlab's call & Boundary conditions \\
			\hline
			\hline
			\begin{minipage}{4cm}
				\includegraphics[width=\linewidth]{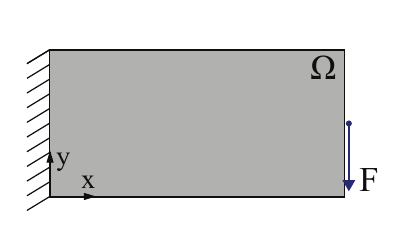}
			\end{minipage}
			&
			{\begin{lstlisting}[numbers=none,breaklines=true,gobble=3]
			UNVARTOP_2D_compliance (100, 50, 12, 0, 0.65, 0, 0.5)\end{lstlisting}} & 
			{\begin{lstlisting}[numbers=none,breaklines=true,gobble=3,belowskip=0pt]
			F(n_unkn*find(coord(:,2)==round(0.5*nely) & coord(:,1)==nelx),1) = -0.01*nelx;
			fixed_dofs = reshape(n_unkn*find(coord(:,1)==0)+(-n_unkn+1:0),1,[]);
			active_node = []; passive_node = [];\end{lstlisting}} \\
			\hline
			\begin{minipage}{4cm}
				\includegraphics[width=\linewidth]{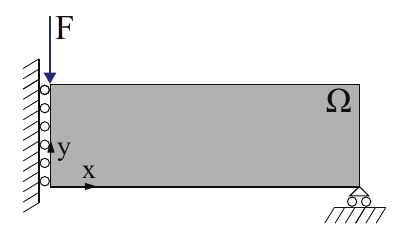}
			\end{minipage}
			& {\begin{lstlisting}[numbers=none,breaklines=true,gobble=3]
			UNVARTOP_2D_compliance (150, 50, 10, 0, 0.6, 0, 1)\end{lstlisting}} &
			{\begin{lstlisting}[numbers=none,breaklines=true,gobble=3,belowskip=0pt]
			F(n_unkn*find(coord(:,2)==nely & coord(:,1)==0),1) = -0.01*nelx;
			fixed_dofs = [reshape(n_unkn*find(coord(:,1)==0)-1,1,[]),...
						  reshape(n_unkn*find(coord(:,1)==nelx & coord(:,2)==0),1,[])];
			active_node = []; passive_node = [];\end{lstlisting}} \\
			\hline
			\begin{minipage}{4cm}
				\includegraphics[width=\linewidth]{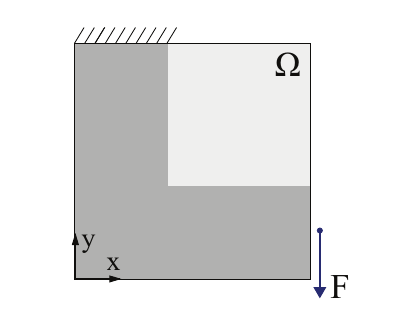}
			\end{minipage}
			&
			{\begin{lstlisting}[numbers=none,breaklines=true,gobble=3]
			UNVARTOP_2D_compliance (100, 100, 12, 0.36, 0.75, 0, 0.5)\end{lstlisting}} & 
			{\begin{lstlisting}[numbers=none,breaklines=true,gobble=3,belowskip=0pt]
			F(n_unkn*find(coord(:,2)==round(0.2*nely) & coord(:,1)==nelx),1) = -0.01*nelx;
			fixed_dofs = reshape(n_unkn*find(coord(:,1)<=0.4*nelx & coord(:,2)==nely)+(-n_unkn+1:0),1,[]);
			active_node = [];
			passive_node = find(coord(:,1)>ceil(nelx*0.4) & coord(:,2)>ceil(nely*0.4));\end{lstlisting}} \\
			\hline
			\begin{minipage}{4cm}
				\includegraphics[width=\linewidth]{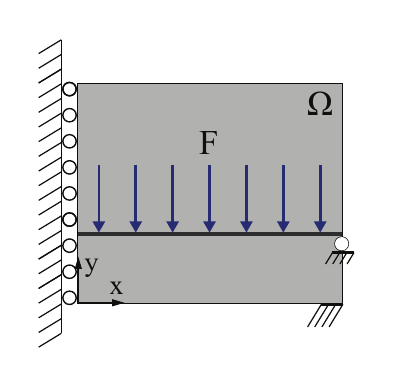}
			\end{minipage}
			 &
			{\begin{lstlisting}[numbers=none,breaklines=true,gobble=3]
			UNVARTOP_2D_compliance (240, 200, 32, 0, 0.775, 0, 0.5)\end{lstlisting}} &
			{\begin{lstlisting}[numbers=none,breaklines=true,gobble=3,belowskip=0pt]
			F(n_unkn*find(coord(:,2)==floor(nely*1.6/5)),1) = -0.01*nelx;
			fixed_dofs = [reshape(n_unkn*find(coord(:,1)==0)-1,1,[]),...
						  reshape(n_unkn*find(coord(:,1)>=5.75/6*nelx & coord(:,2)==0)+(-n_unkn+1:0),1,[]),...
						  reshape(n_unkn*find(coord(:,1)==nelx & coord(:,2)==floor(nely*1.5/5)),1,[])];
			active_node = find(coord(:,2)>=nely*1.5/5 & coord(:,2)<=nely*1.6/5);
			passive_node = [];\end{lstlisting}}\\
			\hline
			\begin{minipage}{4cm}
				\includegraphics[width=\linewidth]{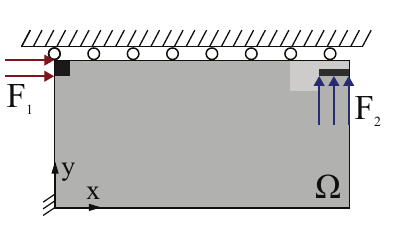}
			\end{minipage} &
			{\begin{lstlisting}[numbers=none,breaklines=true,gobble=3]
			UNVARTOP_2D_ complmechanism (150, 75, 14, 0, 0.85, -2, 0.5)\end{lstlisting}} &
			{\begin{lstlisting}[numbers=none,breaklines=true,gobble=3,belowskip=0pt]
			F(n_unkn*find(coord(:,2)>=0.9*nely & coord(:,1)==0)-1,1) = 0.0001*nelx;
			F(n_unkn*find(coord(:,2)==round(0.9*nely) & coord(:,1)>=0.9*nelx),2) = 0.0001*nelx;
			fixed_dofs = [reshape(n_unkn*find(coord(:,2)==nely),1,[]), reshape(n_unkn*find(coord(:,1)==0 & coord(:,2)<=0.1*nely)+(-n_unkn+1:0),1,[])];
			active_node = [find(coord(:,2)>0.9*nely&coord(:,1)<0.05*nelx); find(coord(:,2)>0.9*nely&coord(:,2)<=0.95*nely&coord(:,1)>=0.9*nelx)];
			passive_node = [find(coord(:,1)>0.8*nelx & coord(:,1)<0.9*nelx & coord(:,2)>0.8*nely); find(coord(:,1)>=0.9*nelx & coord(:,2)>0.95*nely)];\end{lstlisting}} \\
			\hline
		\end{supertabular}
	\end{table*}

	\begin{table*}
		\begin{supertabular}{ ||c m{4cm} m{8cm}|| }
			\hline
			\hline
			Domain & Matlab's call & Boundary conditions \\
			\hline
			\hline
			\begin{minipage}{4cm}
				\includegraphics[width=\linewidth]{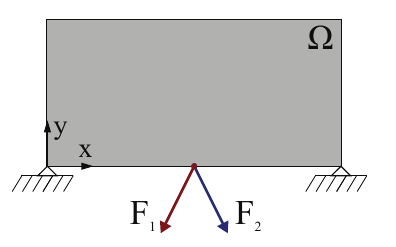}
			\end{minipage}
			&
			{\begin{lstlisting}[numbers=none,breaklines=true,gobble=3]
			UNVARTOP_2D_multiload (200, 100, 24, 0, 0.6, 0, 0.5)\end{lstlisting}} &
			{\begin{lstlisting}[numbers=none,breaklines=true,gobble=3,belowskip=0pt]
			F(n_unkn*find(coord(:,2)==0 & coord(:,1)==round(nelx/2))-1,1) = -0.01*nelx;
			F(n_unkn*find(coord(:,2)==0 & coord(:,1)==round(nelx/2)),1) = -2*0.01*nelx;
			F(n_unkn*find(coord(:,2)==0 & coord(:,1)==round(nelx/2))-1,2) = 0.01*nelx;
			F(n_unkn*find(coord(:,2)==0 & coord(:,1)==round(nelx/2)),2) = -2*0.01*nelx;
			fixed_dofs = reshape(n_unkn*find((coord(:,1)==0 & coord(:,2)==0) | (coord(:,1)==nelx & coord(:,2)==0))+(-n_unkn+1:0),1,[]);
			active_node = []; passive_node = [];\end{lstlisting}} \\
			\hline
			\hline
		\end{supertabular}
	\end{table*}

	\subsection{Cantilever beam}
	
		\begin{figure}[pb]
			\includegraphics[width=8cm]{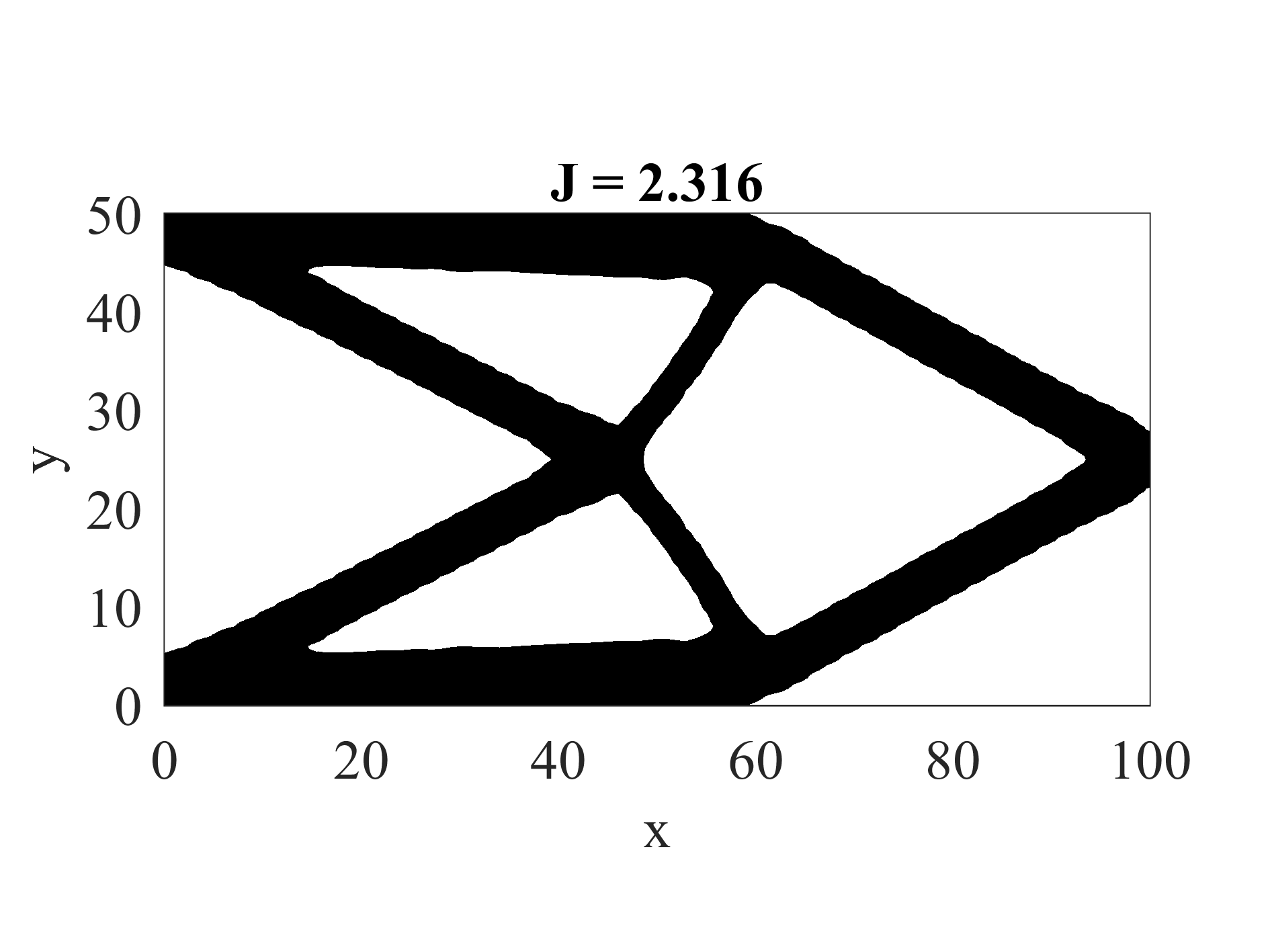}
			\caption{Cantilever beam (load applied at the middle): optimal topology layout.}
			\label{fig_Cantilever_middle_top}
		\end{figure}
		
		\begin{figure}[pt]
			\includegraphics[width=8cm]{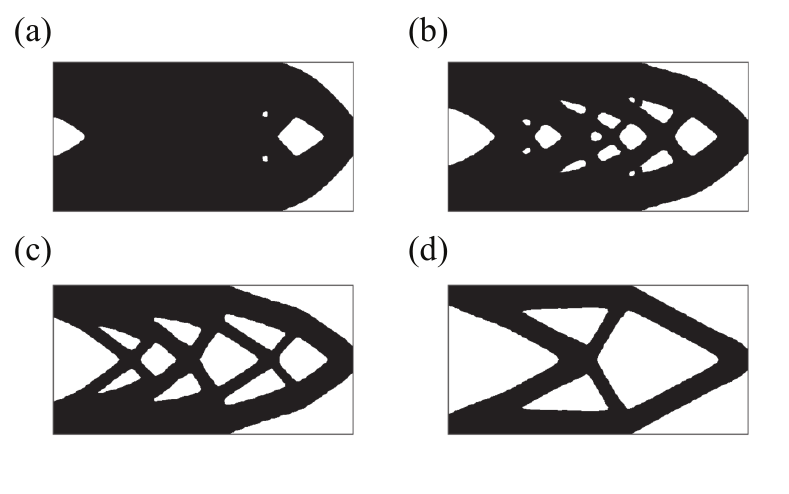}
			\caption{Cantilever beam: topology evolution. (a) optimal topology at $t_{ref}=0.11$, (b) optimal topology at $t_{ref}=0.22$, (c) optimal topology at $t_{ref}=0.38$ and (d) optimal topology at $t_{ref}=0.54$.}
			\label{fig_Cantilever_middle_top_evolution}
		\end{figure}
		
		A variation of the initial examples is now performed. In this case, the load is not applied at the bottom-right corner but in the middle of the right side of the domain, as depicted in the first row of Table \ref{tab_example}. Dirichlet conditions are not modified, i.e. the displacements are prescribed on the left boundary of the domain. The optimal topology layout, for the last time-step, is illustrated in Figure \ref{fig_Cantilever_middle_top}, with the values from Table \ref{tab_example}. That is, the interval of interest $[0,0.65]$ is discretized with 12 equally spaced steps and the regularization parameter is prescribed to $\tau=0.5$. In addition, the topology evolution, shown in the animation (\href{https://github.com/DanielYago/UNVARTOP/blob/master/Online_Resources/ESM_04.gif}{Online Resource 4}), is displayed in Figure \ref{fig_Cantilever_middle_top_evolution} for time-steps 2, 4, 7, and 10. Similar results obtained with other optimization techniques can be found in \cite{Suresh2010,Biyikli2015,Zhang2015,Da2017}.
			
	\subsection{Messerschmitt-Bölkow-Blohm (MBB) beam}
		
\begin{figure}[pt]
			\includegraphics[width=8cm]{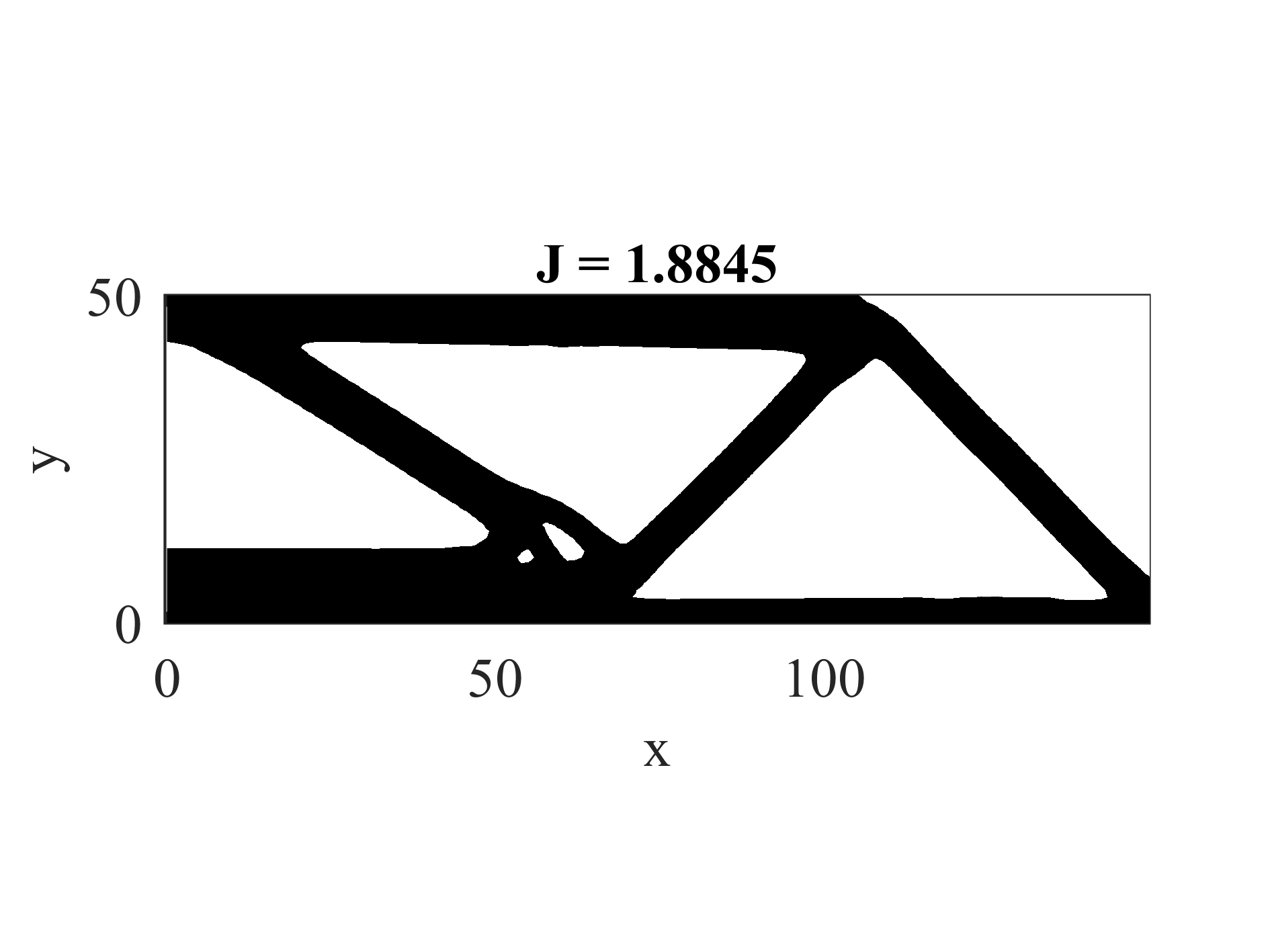}
			\caption{MBB beam: optimal topology layout.}
			\label{fig_MBB_top}
		\end{figure}

		Half of the MBB-beam, with an aspect ratio of 3:1, is optimize in the second example. Symmetry is assumed on the left side of the domain and the vertical displacement at the bottom-right corner is constrained, as observed in Table \ref{tab_example}, and only a point-wise load is applied at the top-left corner. The last requested $t_{ref}=0.6$ is achieved in 10 time steps. Figure \ref{fig_MBB_top} depicts the resulting optimal topology layout using the provided code, adapted with the corresponding boundary conditions (see second row of Table \ref{tab_example}). Additionally,  \href{https://github.com/DanielYago/UNVARTOP/blob/master/Online_Resources/ESM_05.gif}{Online Resource 5} displays the animation of optimal topologies for the given time-steps. The results are comparable to those presented by \cite{Sigmund2001,Andreassen2010,Tavakoli2013,Zhang2015,Da2017}, among others.
		
	\subsection{L-Shape structure}
	
\begin{figure}[pt]
			\includegraphics[width=8cm]{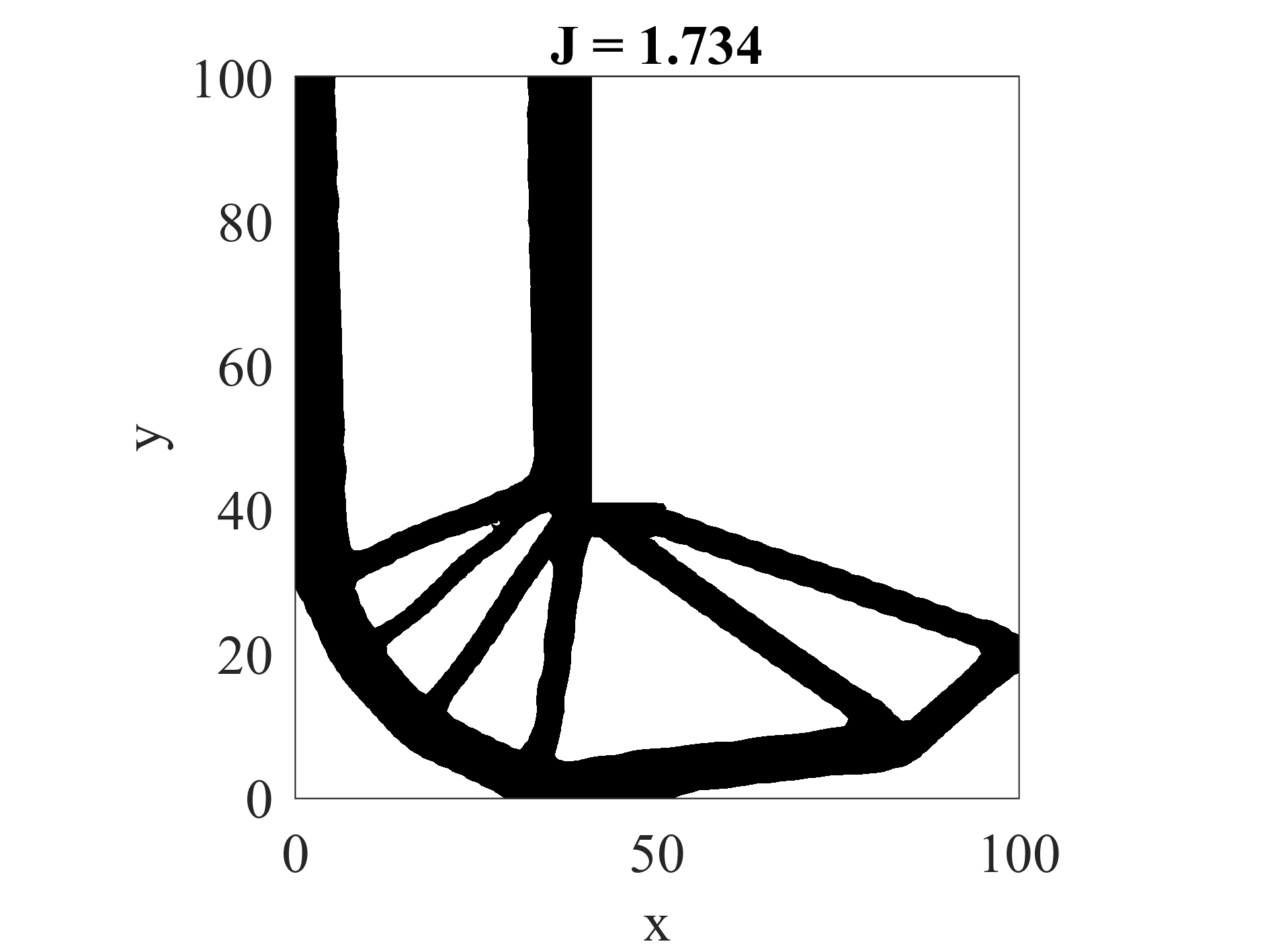}
			\caption{L-shape structure: optimal topology layout.}
			\label{fig_Lshape_top}
		\end{figure}

		The L-Shape structure, shown in Table \ref{tab_example}, represents a simplified version of a hook. The domain has a prescribed void zone in the top right area, defined by $x_i\ge0.4$ and $y_i\ge0.4$. All the nodes contained in this area are listed in \lstinline|passive_node|. A single vertical load is applied on the right side of the domain at $y=0.2$.\footnote{All measures are relative to the dimensions of the domain.} The nodes on the top-left boundary ($y=1$ and $x<0.4$) are fixed. The optimal configuration, shown in Figure \ref{fig_Lshape_top}, is obtained by the inputs described in Table \ref{tab_example}. As in previous examples, the topology evolution is animated in  \href{https://github.com/DanielYago/UNVARTOP/blob/master/Online_Resources/ESM_06.gif}{Online Resource 6}. Similar optimal designs are obtained by \citet{Biyikli2015} and \citet{Liu2014}, for 2D and 3D problems, respectively.
		
	\subsection{Bridge}
	
		The fourth numerical example in Table \ref{tab_example} corresponds to a bridge, which domain is given by 12x5 rectangle. However, only half of it is optimize thanks to the central symmetry. Then, the horizontal displacement on the left side of the domain is prescribed to 0. In addition, the domain is supported by a small segment on the bottom-right corner of it and the vertical displacement is prescribed at the right side of the road. A distributed vertical downside load is applied on the road, which does not change throughout the optimization procedure (i.e. it can not be removed since all its nodes are included in \lstinline|active_node| list). The corresponding boundary conditions of this problem are listed in Table \ref{tab_example}. 
		
		\begin{figure}[pb]
			\includegraphics[width=8cm]{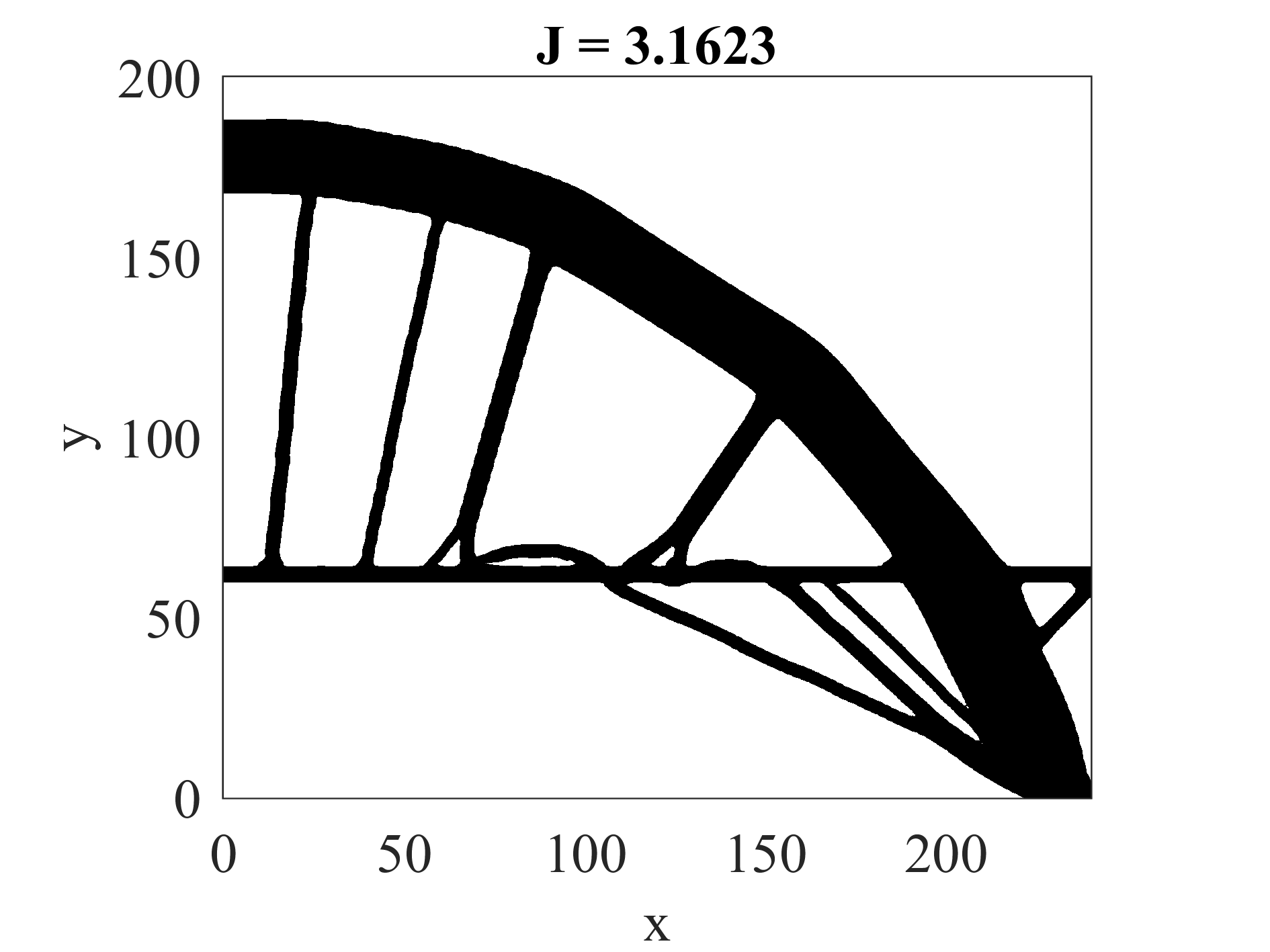}
			\caption{Bridge: optimal topology layout.}
			\label{fig_Bridge_top}
		\end{figure}
		
		The optimal topology, at $t_{ref}=0.775$, is displayed in Figure \ref{fig_Bridge_top}, along with the corresponding animation in  \href{https://github.com/DanielYago/UNVARTOP/blob/master/Online_Resources/ESM_07.gif}{Online Resource 7}. The topology in Figure \ref{fig_Bridge_top} is closely similar to that obtained by \citet{Feijoo2005} and \citet{Liang2002}. Furthermore, the design can be compared with the solution of a multi-load problem done by \cite{Lopes2015}.
					
	\subsection{Gripper mechanism}
	
		\begin{figure}[pb]
			\includegraphics[width=8cm]{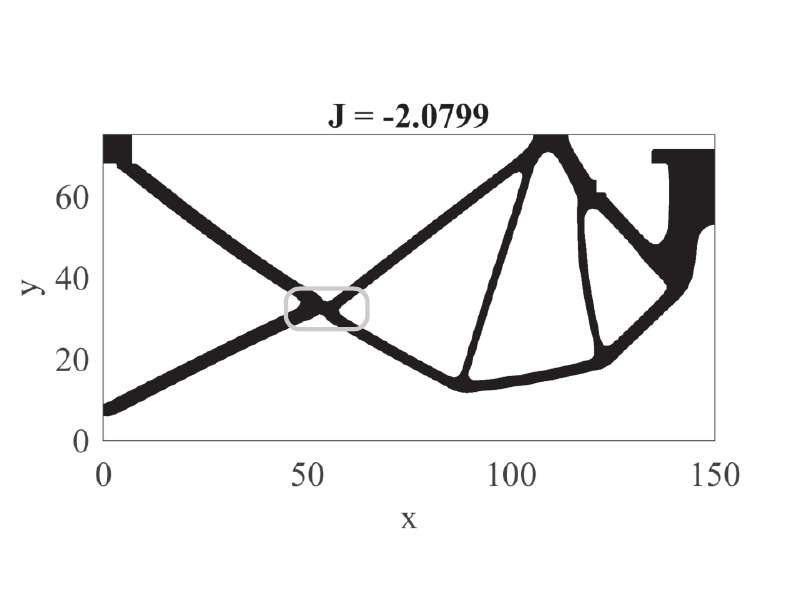}
			\caption{Gripper (compliant mechanism): optimal topology layout. The central hinge is highlighted with a gray square.}
			\label{fig_Gripper_top}
		\end{figure}
		
		\begin{figure}[pb]
			\includegraphics[width=8cm]{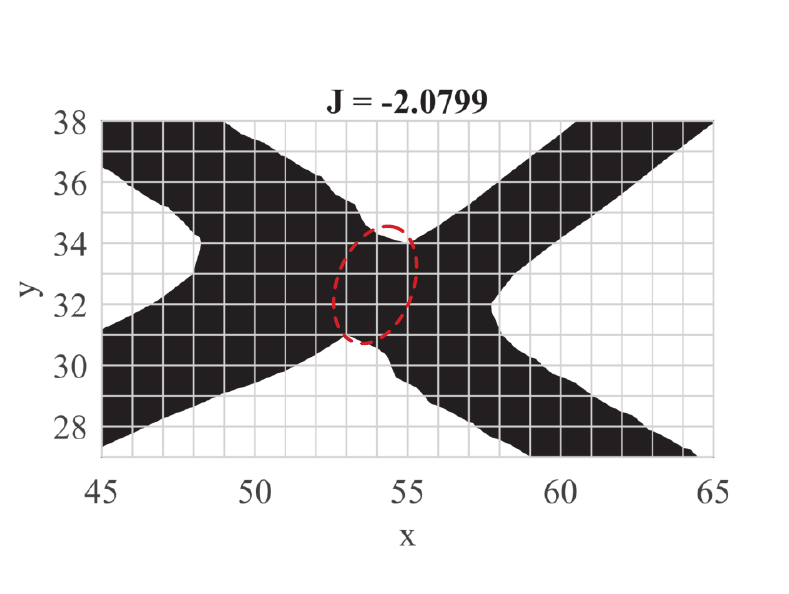}
			\caption{Gripper (compliant mechanism): close-up view of the central hinge.}
			\label{fig_Gripper_top_hinge}
		\end{figure}
		
		Let us now consider a compliant mechanism different from the one explained in section \ref{sec_code_complmechanism} and inspired by \cite{Oliver2019,Yamada2010}. The goal of this optimization is to maximize the compressive displacement at the \emph{output port} (vertical displacement at the top-right side) when an horizontal force is applied at the \emph{input port} (top-left side of the domain). The domain is supported by a small area in the bottom-left corner and symmetry is applied on the top side of $\Omega$, as it can be observed in Table \ref{tab_example} (fifth row). A small area in the \emph{output port} is set to soft material (i.e. included in \lstinline|passive_node| list) in order to represent the gap in the jaws of the gripper. Furthermore, some stiff material areas are restricted in both ports, and the corresponding spring stiffness values must be replaced by $0.01$. 
		
		The \emph{pseudo-time} is updated following an exponential expression in 14 time steps. The given optimal topology of Figure \ref{fig_Gripper_top} is obtained evoking \lstinline|UNVARTOP_2D_compl mechanism| function with the appropriate boundary conditions. A close-up view of the central hinge is illustrated in Figure \ref{fig_Gripper_top_hinge}, where the flexible (thin) material, circled in red, performs as a hinge. The compliant mechanism of Figure \ref{fig_Gripper_top} is animated in  \href{https://github.com/DanielYago/UNVARTOP/blob/master/Online_Resources/ESM_08.gif}{Online Resource 8}, where the displacements are updated following a sinus function. 

	\subsection{Michell multi-load structure}
	
		\begin{figure}[pb]
			\includegraphics[width=8cm]{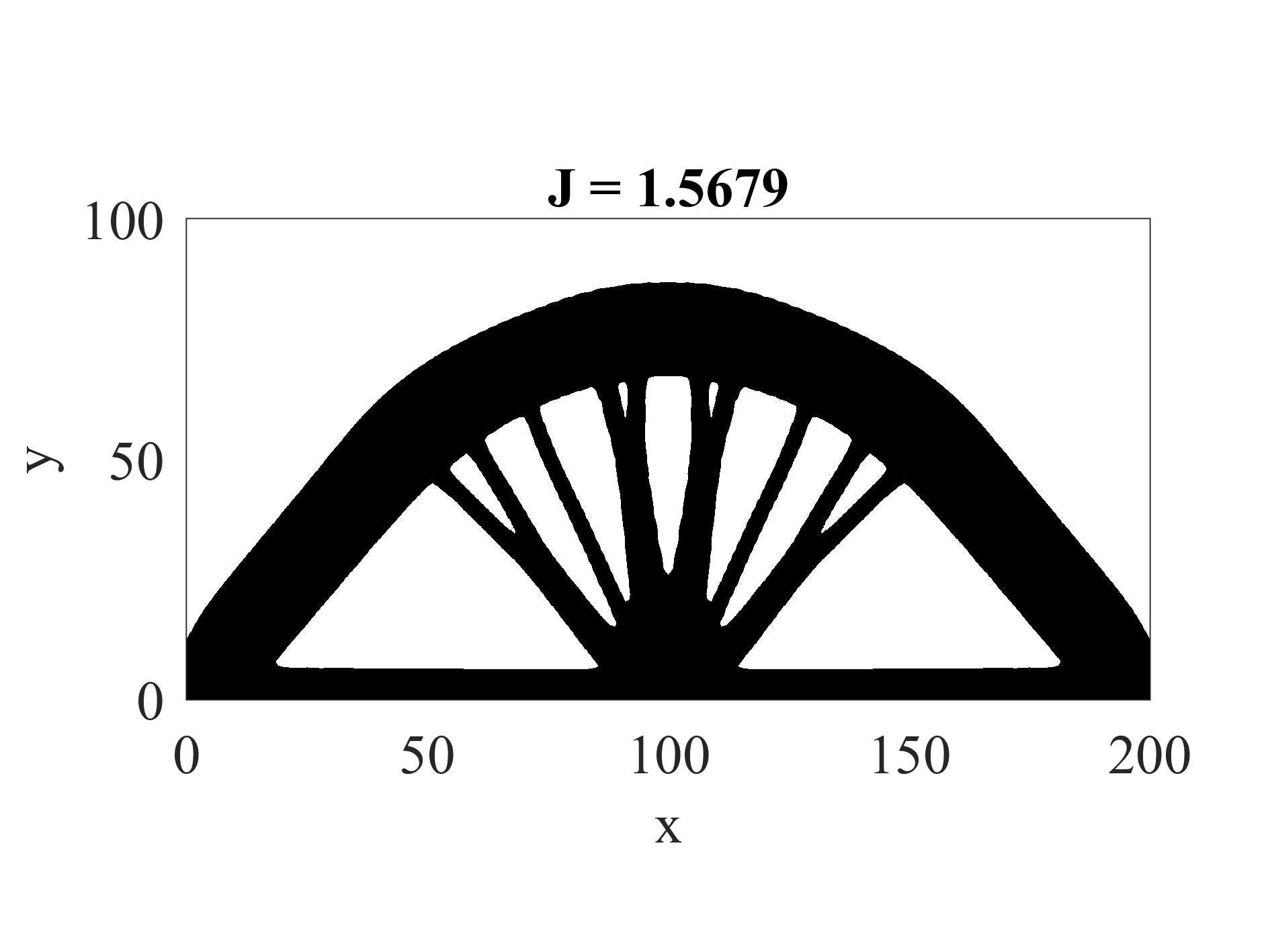}
			\caption{Multi-load michell structure: optimal topology layout.}
			\label{fig_Michell_top}
		\end{figure}
		
		\begin{figure}[pb]
			\includegraphics[width=8cm]{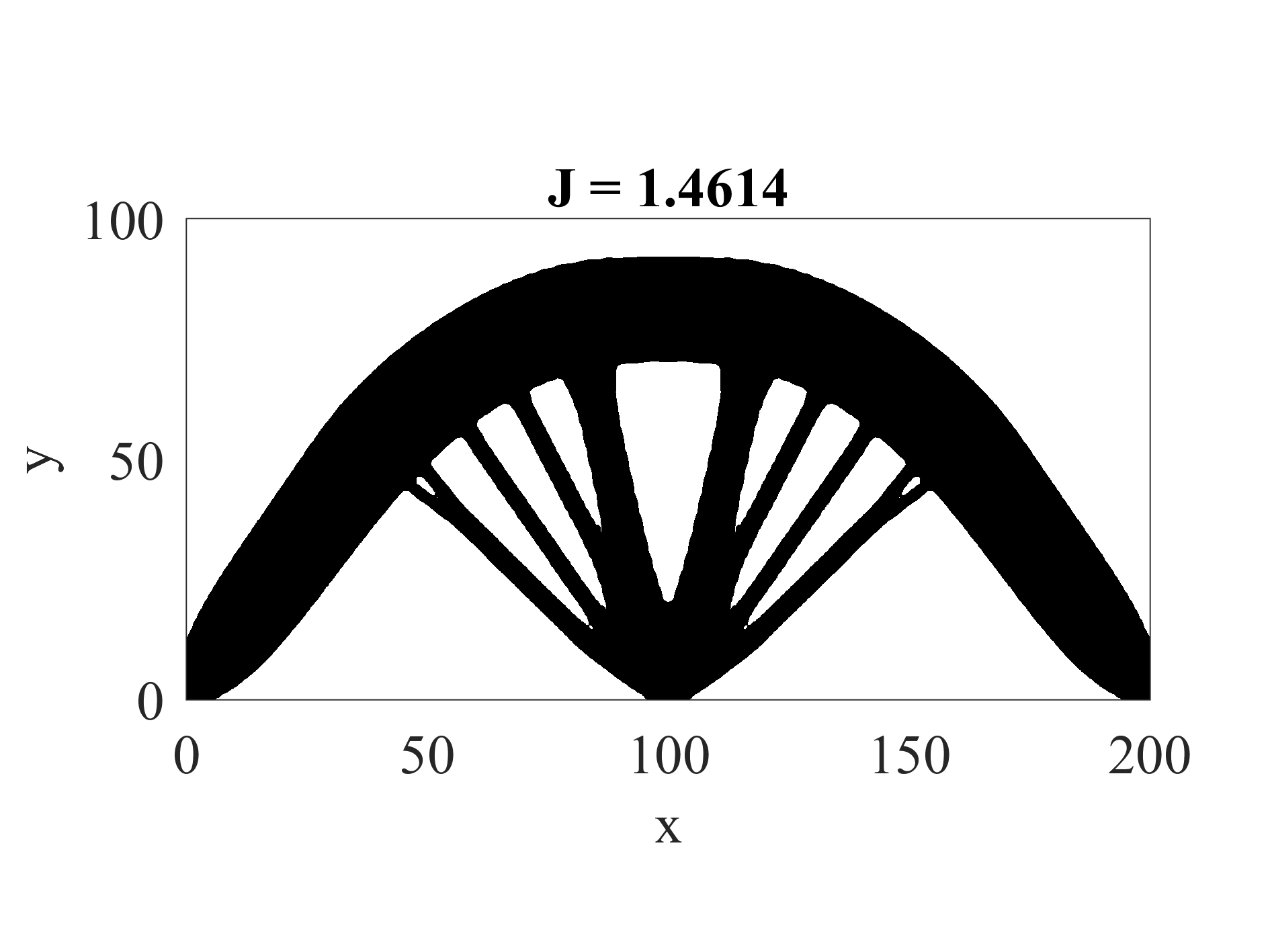}
			\caption{Multi-load michell structure: optimal topology layout when loads are applied at the same time.}
			\label{fig_Michell_top_single}
		\end{figure}
		
		The last numerical example corresponds to a multi-load mean compliance problem with two loading states. The 2x1 rectangular domain is supported by its two bottom corners and subjected to a pair of forces in the middle of the bottom side at an angle of 30° with respect to the vertical. The desired \emph{pseudo-time} is prescribed to $0.6$, which optimal topology is illustrated in Figure \ref{fig_Michell_top}. The topology animation is given in  \href{https://github.com/DanielYago/UNVARTOP/blob/master/Online_Resources/ESM_09.gif}{Online Resource 9}. 
		
		The topology layout, as already noted, deviates from the corresponding optimal topology when both loads are applied at the same time, as shown in Figure \ref{fig_Michell_top_single}. In this setting, the bottom bars, which connect the supporting nodes with the central node, have been removed. The problem definition is based on \cite{Lopes2015}. Other variations can be found in \cite{Challis2009,Tavakoli2013}.

\section{Extensions} \label{sec_extension}

	\subsection{Bisection algorithm}
	
		The bisection algorithm of the \emph{cutting\emph{\&}bisection} algorithm, which estimates the solution as the midpoint of the bracketing interval (see section \ref{sec_code_bisection}), can be easily improved by introducing either a \emph{regula falsi method} \cite{R.Bulirsch2010} or a more sophisticated method, like the \emph{Anderson-Björk with Illinois algorithm} \cite{Anderson1973}. These two mathematical techniques reduce the number of iterations required to find the root of the \emph{constraint equation} ((\ref{eq_topopt_prob_compliance})-b), $\cal{C}$.
		
		\subsubsection{Regula falsi}
			
			In order to compute the test \lstinline|lambda|\footnote{The \emph{Lagrange multiplier} is denoted as \lstinline|lambda| in the code.} through the \emph{regula falsi} approximation inside the \emph{bisection algorithm} \cite{R.Bulirsch2010}, line 143 must be replaced by 
\begin{lstlisting}[language=Matlab,firstline=159,lastline=159,firstnumber=143]
	lambda = l1 - c1*(l2-l1)/(c2-c1);
\end{lstlisting}
			where \lstinline|l1| and \lstinline|l2| stand for the left and right $\lambda$ brackets, while \lstinline|c1| and \lstinline|c2| are respectively the corresponding constraint values. The linear interpolation with the endpoints of the bracketing interval is used to find the value of the root, i.e. the root is approximated as the intersecting point between the line joining the extremes and the x-axis. Next, the subinterval is updated by checking the sign of the constraint equation at \lstinline|lambda|, as mentioned in section \ref{sec_code_bisection}, until the tolerance is attained. 
		
		\subsubsection{Anderson-Björck with Illinois algorithm}
		
			The \emph{regula falsi} method usually converges faster than the the regular \emph{bisection algorithm}. However, for some specific situations, it can show slower convergence. To avoid these numerical instabilities, the \emph{Anderson-Björck algorithm with an Illinois algorithm} \cite{Anderson1973} is implemented. \footnote{On one hand, the \emph{Illinois} method \cite{Dowell1971} seeks to eliminate the ill-condition generated by permanently retaining one of the end-points (always set to the left bracket in the code). This issue is fixed by multiplying the retained extreme point by $g=0.5$. On the other hand, \emph{Anderson-Björck algorithm} improves the \emph{regula falsi approach} by combining linear interpolation (when the left bracket should be replaced) with parabolic interpolation (when the right bracket should be replaced). Furthermore, it includes an \emph{Illinois-scheme} with $g=1-\frac{{\cal{C}}(\lambda)}{{\cal{C}}(\lambda_2)}$, when $g$ is positive, or $0.5$, otherwise.} Lines 143 and 152 are changed to:
\begin{lstlisting}[language=Matlab,firstline=159,lastline=159,firstnumber=143]
	lambda = l1 - c1*(l2-l1)/(c2-c1);
\end{lstlisting}
			and
\begin{lstlisting}[language=Matlab,firstline=171,lastline=172,numberstyle=\tiny\color{codegray}152-]
	if c2*Tol_constr<=0; l1=l2; c1=c2; l2=lambda; c2=Tol_constr;
	else; g=1-Tol_constr/c2; g=(g-0.5)*(g>0)+0.5; l2=lambda; c1=g*c1; c2=Tol_constr; end
\end{lstlisting}
			With these changes, the number of iterations and the computational cost/time of the optimization procedure is reduced.			
			
	\subsection{Plane-strain assumption}
	
		The corresponding constitutive tensor, $\overline{\mathbb{C}}$, for the plane-strain assumption
		\begin{equation}
			\overline{\mathbb{C}}^{Pstrain}=\dfrac{E}{(1-\nu)(1-2\nu)}\left[  
			\begin{array}{*3{C{3em}}}
				1-\nu & \nu & 0 \\
				\nu & 1-\nu & 0  \\
				0   & 0     & \dfrac{1-2\nu}{2} \\
			\end{array}
			\right]
		\end{equation}
		can be easily used by replacing the definition of the constitutive tensor of the plane-stress assumption, see section \ref{sec_thry_state_eq}, in lines 126-1287 with the following 
\begin{lstlisting}[language=Matlab,firstline=134,lastline=135,firstnumber=126]
function [DE] = D_matrix_strain(E,nu) %Planestrain
DE = E/((1+nu)*(1-2*nu))*[(1-nu) nu 0;nu (1-nu) 0;0 0 (1-2*nu)/2];
\end{lstlisting}
		Line 27 must be also modified, to call the \lstinline{D_matrix_strain} function, to
\begin{lstlisting}[language=Matlab,firstline=34,lastline=34,firstnumber=27]
[DE] = D_matrix_stress(E0,nu);
\end{lstlisting}
		
	\subsection{Augmented Lagrangian to impose volume constraint}
	
		The constraint equation (\ref{eq_topopt_prob_compliance})-b, ${\cal{C}}$, can be also imposed through an Augmented Lagrangian method \cite{Luenberger2016}, which updates the lagrangian multiplier according the following definition
		\begin{equation} \label{eq_lambda_aug}
			\lambda_{i+1}=\lambda_{i} + \rho {\cal{C}}_{i} \mcolon
		\end{equation}
		to prescribed an equality constraint. The penalty value, $\rho$, can be either set to a constant value or increased along iterations, which improves convergence rate. Then, the penalty coefficient is updated as
		\begin{equation} \label{eq_penalty_aug}
			\rho_{i+1} = \left\{ 
			\begin{split}
				&\min\left( 1.02 \rho_i , 100 \rho_0 \right) &&\text{for} \quad |{\cal{C}}_{i+1}-{\cal{C}}_{i}|<10^{-3} \\
				&\rho_i &&\text{otherwise} \mcolon
			\end{split}
			\right.
		\end{equation}
		where $\rho_0$ corresponds to the initial penalty value and $i$ represents the i-th iteration. The values $1.02$ and $100$ can be modified at the user's discretion, and will highly depend on each specific numerical example. 
		In this implementation, the Lagrangian equation (\ref{eq_Lagrangian}) is defined as
		\begin{equation} \label{eq_Lagrangian_aug}
			{\cal L} = J + \lambda {\cal{C}} + \dfrac{1}{2} \rho {\cal{C}}^2 \mdot
		\end{equation}
		
		In order to impose the constraint with this methodology, a few changes need to be made to the original code of Appendix \ref{app_matlab_code}. First, the initialization of constraint vector and the penalty value must be initialized by inserting 
\begin{lstlisting}[language=Matlab,firstline=65,lastline=65,numbers=none]
Tol_constr_vec = []; rho = rho0;
\end{lstlisting}
		between lines 57 and 58, and the constraint must be computed before starting the optimization loop, just below line 64:
\begin{lstlisting}[language=Matlab,firstline=73,lastline=73,numbers=none]
	Tol_constr = t_ref - vol;
\end{lstlisting}
		and stored in the corresponding vector after line 95
\begin{lstlisting}[language=Matlab,firstline=113,lastline=113,numbers=none]
		Tol_constr_vec = [Tol_constr_vec,abs(Tol_constr)];
\end{lstlisting}
		The convergence criteria of line 65 must be also changed to include the constraint equation as an extra convergence condition by introducing \lstinline{abs(Tol_constr)>1e-3}.
		
		Second, the \emph{bisection algorithm} (lines 137-146) and its function call in the optimization loop (line 88) must be replaced with the corresponding updating of $\lambda$ and $\rho$ (equations (\ref{eq_lambda_aug}) and (\ref{eq_penalty_aug})), defined as
\begin{lstlisting}[language=Matlab,firstline=154,lastline=159,numbers=none]
function [lambda,rho,chi,psi,vol,Tol_constr] = find_volume (iter,xi,connect,active_node,passive_node,t_ref,lambda,rho,rho0,alpha0,Tol_constr,Tol_constr_vec)
lambda = lambda + rho * Tol_constr;
psi = xi - lambda; psi(passive_node) = -alpha0; psi(active_node) = alpha0;
[vol,chi] = compute_volume (psi,connect);
Tol_constr = -(vol-t_ref);
if iter>=3; rho = min(0.02*rho*(abs(diff(Tol_constr_vec(end-1:end)))<1e-3) + rho,100*rho0); end
\end{lstlisting}		
		and
\begin{lstlisting}[language=Matlab,firstline=105,lastline=105,firstnumber=88]
		[lambda,rho,chi_n,psi,vol,Tol_constr] = find_volume (iter,xi,connect,active_node,passive_node,t_ref,lambda,rho,rho0,alpha0,Tol_constr,Tol_constr_vec);
\end{lstlisting}
		
		Last, the cost function must be computed according equation (\ref{eq_Lagrangian_aug}), which takes into account the constraint equation. The additional terms are summed in one extra line under line 69:
\begin{lstlisting}[language=Matlab,firstline=103,lastline=103,numbers=none]
		J = J + nelx*nely*(lambda*Tol_constr + rho*Tol_constr^2)/J_ref*xi_norm;
\end{lstlisting}
		
	\subsection{Thermal problem} \label{sec_ext_thermal}
	
		According to \citet{Yago2020}, the implementation of the thermal compliance problem is rather analogous to the structural mean compliance problem, detailed in section \ref{sec_thry_compliance}. In that case, the temperature, $\hat{\theta}$, is the only unknown per node (\lstinline|n_unkn=1|) and the steady-state problem is used as the state equation. Therefore, the cost function (\ref{eq_compliance_cost_function}) has to be replaced by
		\begin{equation} \label{eq_thermal_compliance_cost_function}
			\begin{split}
				{\cal J}&(\theta_\chi)\equiv \frac{1}{2} l(\theta_{\chi})=\frac{1}{2} a_{\chi}(\theta_{\chi},\theta_\chi)=\\
					&\equiv \frac{1}{2} \left(\int_{\Omega} \bm{\nabla}\theta_{\chi} \cdot{\Kappalarge}_{\chi}\cdot\bm{\nabla}\theta_{\chi}\domega\right)=\int_\Omega {\cal U_\chi}\domega \mcolon
			\end{split}
		\end{equation}
		where $l(\theta_{\chi})$ and $a_{\chi}(\theta_{\chi},\theta_\chi)$ correspond to the bilinear forms of the thermal problem. Furthermore, $\bm{\nabla}\theta_{\chi}$ and ${\Kappalarge}_{\chi}$ represent the thermal gradient vector and the symmetric second order thermal conductivity tensor, respectively. Unlike the elastic material behavior used in section \ref{sec_thry_state_eq}, the conductive material follows Fourier's law, i.e. the heat flux is proportional to the thermal gradient by $\bm{q}\bbxchi = - \Kappalarge\bbxchi\cdot\bm{\nabla}\theta_{\chi}\bbx$. 
		
		The state equation (\ref{eq_weak_problem2}) must be also substituted by
		\begin{empheq}[left=\empheqlbrack,right=\hspace{-0.1cm}]{align}
			&\text{Find the temperature field  ${\pmb \theta}_\chi\in{\cal{U}}(\Omega)$ such that} \notag \\
			& \hspace{0.25cm} a(w,\theta_\chi) = l(w) \quad \forall w\in {\cal V}(\Omega) \label{eq_weak_problem3}\\
		    &\text{where} \notag \\
			&\hspace{0.25cm}a(w,\theta_\chi) = \int_{\Omega}{\bm{\nabla} w\bbx\cdot\Kappalarge_\chi\bbx\cdot\bm{\nabla}\theta_\chi\bbx}\domega \mcolon \label{eq_lhs_thermal_problem} \\
			&\hspace{0.25cm}l(w) = - \int_{\partial_{q}\Omega}{w\bbx{\overline{q}}\bbx}\dgamma  \notag \\
			&\hspace{1.25cm}+ \int_{\Omega}{w\bbx \hS_\chi\bbx}\domega \mcolon \label{eq_rhs_thermal_problem}
		\end{empheq}
		where ${\cal{U}}(\Omega)$ and ${\cal V}(\Omega)$ stand for the corresponding set of admissible temperature fields and the corresponding space of admissible virtual temperature fields, respectively.  $\hS\bbxchi$ and ${\overline{q}}\bbx$ correspond respectively to the heat source function and the prescribed heat flux on the boundaries of $\Omega$.
		
		After applying the RTD to equation (\ref{eq_thermal_compliance_cost_function}), mimicking the procedure described in section \ref{sec_thry_compliance}, the resultant \emph{pseudo-energy}, $\xi\bbxhat$, is expressed as
		\begin{equation} \label{eq_topopt_xi_thermalcompliance}
			\xi\bbxhatchi = 2m_{\kappa} \left(\chi_{\kappa}\bbxhat\right) ^{m_{\kappa}-1}{\overline{\cal U}}\bbxhat{{\Delta\chi}_{\kappa}\bbxhat } \mcolon
		\end{equation}
		with
		\begin{equation}
			\overline{\cal U}\bbxhat=\dfrac{1}{2}\left( \bm{\nabla}\theta_{\chi} \cdot {\overline{\Kappalarge}} \cdot \bm{\nabla}\theta_{\chi}\right)\bbxhat \ge 0 \mdot
		\end{equation}	
		
		Several modification to the provided code, based on equation (\ref{eq_thermal_compliance_cost_function}) to (\ref{eq_topopt_xi_thermalcompliance}), are required in order to solve thermal problems. The most relevant ones are listed next: the number of unknowns per node must be set to 1, the gradient matrix, $\mathbf{B}$, must be adjusted to be equal to the Cartesian derivatives, the material property is now the conductivity value of the high conductive material instead of $E$ and $\nu$ and the constitutive tensor $\Kappalarge$ is now defined as
		\begin{equation}
			\Kappalarge = \kappa \left[ 
			\begin{array}{cc}
				K_{11} & K_{12} \\
				K_{21} & K_{22}
			\end{array} \right] \mdot
		\end{equation} 
		In addition, boundary condition must be defined accordingly to the thermal problem.
		
	\subsection{3D extension} \label{sec_ext_3Dprob}
	
		The topology optimization code \emph{\codename} can be readily extended to solve 3D problems. All the functions related to FE analysis must be rewritten, starting from the mesh, the shape matrices $\mathbf{N}$, the corresponding strain-displacement matrices $\mathbf{B}$ and the constitutive tensor $\mathbb{C}$. Therefore, element stiffness matrices should be recomputed, along with the stiffness and mass matrices for the Laplacian regularization. It is recommended to use an iterative solver (e.g. \lstinline|minres| solver) to compute the displacements, as employed for the Laplacian regularization (see section \ref{sec_lapl_reg_code}), in order to reduce computational cost. Function \lstinline|compute_volume| must be slightly adapted to hexahedral elements.
		In addition, functions \lstinline{isosurface}, \lstinline{isocaps} and \lstinline{isonormals} must be used to represent the optimal topology.
		
		It is important to notice that the algorithm inside the topology optimization does not require any modification.
		
\section{Conclusions} \label{sec_conclusions}

	This paper has presented the 2D implementation in Matlab of the unsmooth variational topology optimization approach, previously formulated for structural \cite{Oliver2019} and thermal \cite{Yago2020} topology optimization problems. The paper described and implemented the approach for educational purposes while demonstrating its capabilities and maintaining high computational efficiency and readability of the code. Furthermore, the implementation preserves the finite element analysis of the domain, thus introducing students to the numerical analysis as well as the topology optimization field.
		
	The numerical examples performed in this work illustrate the potential and effectiveness of the technique to tackle a large set of different problems with a volume constraint, e.g minimum mean compliance problems (section \ref{sec_thry_compliance}), multi-load mean compliance problems (section \ref{sec_thry_multicompliance}) and compliant mechanisms synthesis (section \ref{sec_thry_mechanism}). The set of numerical examples include a variety of boundary conditions, active and passive nodes, number of time-steps, along others. Additionally, section \ref{sec_ext_thermal} shows how to easily switch from the structural problem of minimum mean compliance to the thermal problem, where thermal compliance is minimized. Finally, section \ref{sec_ext_3Dprob} provides some guidelines for the extension of the code to the resolution of 3D problems.
	
	The topologies obtained for these examples are comparable to those shown by other researchers using more established techniques (e.g. SIMP method or Level-set method). In addition, smooth topology configuration have been obtained in all the benchmarks with a relatively small number of iterations. That is a feature to be highlighted against more conventional techniques based on elemental densities, such as SIMP method.
	
	In conclusion, the dissemination of this code will provide newcomers in this field a better understanding in how this new topology optimization approach works as well as to encourage future research of this technique for miscellaneous applications.

	The Matlab code, detailed in appendix \ref{app_matlab_code}, along with some variations of it, can be downloaded from the author's GitHub repository \href{https://github.com/DanielYago/UNVARTOP}{https://github.com/DanielYago}. Additional online resources, such as figures and animations, are also stored in the repository.

\begin{acknowledgements}
This research has received funding from the European Research Council (ERC) under the European Union’s Horizon 2020 research and innovation programme (Proof of Concept Grant agreement n 874481) through the project “Computational design and prototyping of acoustic metamaterials for target ambient noise reduction” (METACOUSTIC).
The authors also acknowledge financial support from the Spanish Ministry of Economy and Competitiveness, through the research grant DPI2017-85521-P for the project “Computational design of Acoustic and Mechanical Metamaterials” (METAMAT) and through the “Severo Ochoa Programme for Centres of Excellence in R\&D” (CEX2018-000797-S).
D. Yago acknowledges the support received from the Spanish Ministry of Education through the FPU program for PhD grants. 
\end{acknowledgements}

%
\section*{Conflict of interest}
The authors declare that they have no conflict of interest as regards this work.

\begin{appendices}
	\numberwithin{equation}{section}
	
	\section{Matlab code} \label{app_matlab_code}
	
		\begin{strip}
\begin{lstlisting}[caption={\codenamespc code written in Matlab},firstline=12]
function [iter,J] = UNVARTOP_2D_compliance (nelx,nely,nsteps,Vol0,Vol,k,tau)
n_dim = 2; n_unkn = 2; n_nodes = 4; n_gauss = 4; n = (nelx+1)*(nely+1); h_e = 1; alpha0 = 1e-3;
iter_max_step = 20; iter_min_step = 4; iter_max = 500;
opt = struct('Plot_top_iso',1,'Plot_vol_step',1,'EdgeColor','none','solver_Lap','direct');
%% Vector for assembling matrices
[X,Y] = meshgrid(0:nelx,nely:-1:0); coord = [X(:),Y(:)]; clear X Y
nodenrs = reshape(1:n,1+nely,1+nelx);
nodeVec = reshape(nodenrs(1:end-1,1:end-1)+1,nelx*nely,1); clear nodenrs;
connect = nodeVec+[0 nely+[1 0] -1]; clear nodeVec;
%% Loads and boundary setting for Cantilever beam
F = sparse(n_unkn*n,1);
U = zeros(n_unkn*n,1);
F(n_unkn*find(coord(:,2)==0 & coord(:,1)==nelx),1) = -0.01*nelx;
fixed_dofs = reshape(n_unkn*find(coord(:,1)==0)+(-n_unkn+1:0),1,[]);
active_node = []; passive_node = [];
free_dofs = setdiff(1:(n_unkn*n),fixed_dofs);
U(fixed_dofs,:) = 0;
%% Parameter definition
m = 5; E0 = 1; alpha = 1e-6; beta = nthroot(alpha,m); nu = 0.3;
%% Prepare animation
psi_vec = zeros(size(coord,1),nsteps+1);
chi_vec = zeros(size(connect,1),nsteps+1);
U_vec = zeros(n_unkn*size(coord,1),1,nsteps+1);
%% Finite element analysis preparation
[posgp4,W4] = gauss_points(n_gauss);
[posgp1,W1] = gauss_points(1);
[DE] = D_matrix_stress(E0,nu);
KE = zeros(n_nodes*n_unkn,n_nodes*n_unkn);
KE_i = zeros(n_nodes*n_unkn,n_nodes*n_unkn,n_gauss);
for i=1:n_gauss
	[BE,Det_Jacobian] = B_matrix(posgp4(:,i),n_unkn,n_nodes);
	KE_i(:,:,i) = BE'*DE*BE;
	KE = KE + KE_i(:,:,i)*Det_Jacobian*W4(i);
end
[BE_cut,Det_Jacobian_cut] = B_matrix(posgp1,n_unkn,n_nodes);
K_cut = BE_cut'*DE*BE_cut;
KE_cut = K_cut*Det_Jacobian_cut*W1(1);
edofMat = kron(connect,n_unkn*ones(1,n_unkn)) + repmat(1-n_unkn:0,1,n_nodes);
iK = reshape(kron(edofMat,ones(n_nodes*n_unkn,1))',(n_nodes*n_unkn)^2*nelx*nely,1);
jK = reshape(kron(edofMat,ones(1,n_nodes*n_unkn))',(n_nodes*n_unkn)^2*nelx*nely,1);
%% Laplacian filter preparation
KE_Lap = 1/6* [ 4 -1 -2 -1;-1 4 -1 -2;-2 -1 4 -1;-1 -2 -1 4];
ME_Lap = 1/36*[ 4  2  1  2; 2 4  2  1; 1  2 4  2; 2  1  2 4];
KE_Lap = ME_Lap + (tau*h_e).^2*KE_Lap;
i_KF = reshape(kron(connect,ones(n_nodes,1))',n_nodes^2*nelx*nely,1);
j_KF = reshape(kron(connect,ones(1,n_nodes))',n_nodes^2*nelx*nely,1);
s_KF = reshape(KE_Lap(:)*ones(1,nelx*nely),n_nodes^2*nelx*nely,1); clear KE_Lap ME_Lap;
K_Lap = sparse(i_KF,j_KF,s_KF);
if strcmp(opt.solver_Lap,'direct'); LF = chol(K_Lap,'lower'); clear K_Lap i_KF j_KF s_KF;
else; LF = ichol(K_Lap, struct('type','ict','droptol',1e-3,'diagcomp',0.1)); clear i_KF j_KF s_KF; end
i_xi = reshape(connect',n_nodes*nelx*nely,1);
N_T = N_matrix(posgp4).*W4/4;
%% Loop over steps
psi = alpha0*ones(n,1); psi(passive_node) = -alpha0; psi(active_node) = alpha0; psi_vec(:,1)=psi;
[~,chi] = compute_volume (psi,connect); chi0_step = chi; chi_vec(:,1) = chi';
% Initialize variables
iter = 1; J_vec = []; vol_vec = []; lambda_vec = 0; lambda = 0; fhandle6 = [];
[fhandle2,ohandle2] = plot_isosurface([],[],0,psi,coord,connect,1,opt);
for i_step = 1:nsteps
	[t_ref] = set_reference_volume(i_step,Vol0,Vol,nsteps,k);
	% Main loop by steps
	Tol_chi = 1;
	Tol_lambda = 1;
	iter_step = 1;
	while (((Tol_chi>1e-1 || Tol_lambda>1e-1) && iter_step<iter_max_step) || iter_step<=iter_min_step)
		% FE-analysis
		[K] = assmebly_stiff_mat (chi,KE,KE_cut,beta,m,iK,jK,n_unkn,nelx,nely);
		U(free_dofs,:) = K(free_dofs,free_dofs) \ (F(free_dofs,:) - K(free_dofs,fixed_dofs)*U(fixed_dofs,:));
		if iter == 1; U_vec(:,:,1)=U; J_ref = full(abs(sum(sum(F.*U,1),2))); end; J = full(sum(sum(F.*U,1),2))/J_ref;
		% Calculate sensitivities
		Energy = zeros(n_gauss,nelx*nely);
		id = chi==1|chi==0; int_chi = interp_property (m,m-1,beta,chi(id));
		u_e = reshape(U(edofMat(id,:)',1),n_nodes*n_unkn,[]); w_e = u_e;
		for i=1:n_gauss; Energy(i,id) = sum(w_e.*(KE_i(:,:,i)*u_e),1); end
		Energy(:,id) = int_chi.*Energy(:,id);
		id = ~id; int_chi = interp_property (m,m-1,beta,chi(id));
		u_e = reshape(U(edofMat(id,:)',1),n_nodes*n_unkn,[]); w_e = u_e;
		Energy(:,id) = repmat(int_chi.*sum(w_e.*(K_cut*u_e),1),n_gauss,1);
		if iter == 1; xi_shift = min(0,min(Energy(:))); xi_norm = max(range(Energy(:)),max(Energy(:))); end
		% Apply Laplacian regularization
		xi_int = N_T*(Energy-xi_shift*chi)/xi_norm;
		if strcmp(opt.solver_Lap,'direct')
			xi = LF'\(LF\accumarray(i_xi,xi_int(:),[n 1]));
		else
			[xi,flag] = minres(K_Lap,accumarray(i_xi,xi_int(:),[n 1]),1e-6,500,LF,LF'); assert(flag == 0);
		end
		% Compute topology
		[lambda,chi_n,psi,vol] = find_volume (xi,connect,active_node,passive_node,t_ref,lambda,alpha0);
		lambda_vec = [lambda_vec,lambda];
		Tol_lambda = (lambda_vec(iter+1)-lambda_vec(iter))/lambda_vec(iter+1);
		% Plot topology
		[fhandle2,ohandle2] = plot_isosurface(fhandle2,ohandle2,iter,psi,coord,connect,J,opt);
		% Update variables
		Tol_chi = sqrt(sum((chi-chi_n).^2))/sqrt(sum(chi0_step.^2));
		chi = chi_n;
		fprintf(' Step:%5i It.:%5i Obj.:%11.4f Vol.:%7.3f \n',i_step,iter_step,J,vol);
		iter_step = iter_step+1; iter = iter+1;
		drawnow;
	end
	chi0_step = chi;
	if J<10
		[fhandle6,J_vec,vol_vec] = plot_volume_iter(fhandle6,i_step,J_vec,J,vol_vec,vol,opt.Plot_vol_step,6,'Cost function Step','#step','+-b');
		psi_vec(:,i_step+1)=psi; chi_vec(:,i_step+1)=chi'; U_vec(:,:,i_step+1)=U;
	end
	if iter_step >= iter_max_step; warning('VarTopOpt:Max_iter_step','Maximum number of in-step iterations achieved.'); break; end
	if iter > iter_max; warning('VarTopOpt:Max_iter','Maximum number of iterations achieved.'); break; end
end
%% Animation
Topology_evolution(coord,connect,[Vol0,vol_vec],psi_vec,chi_vec,U_vec);
%%%% %%%% %%%% %%%% %%%% %%%% %%%% %%%% %%%%
function [posgp,W] = gauss_points (n_gauss)
if n_gauss==1; s = 0; w = 2; else; s = sqrt(3)/3*[-1 1]; w = [1 1]; end
[s,t] = meshgrid(s,s); posgp = [s(:) t(:)]';
W=w'*w; W=W(:)';
%%%% %%%% %%%% %%%% %%%% %%%% %%%% %%%% %%%%
function [N] = N_matrix(posgp)
N = 0.25*(1+[-1 1 1 -1]'*posgp(1,:)).*(1+[-1 -1 1 1]'*posgp(2,:));
%%%% %%%% %%%% %%%% %%%% %%%% %%%% %%%% %%%%
function [BE,Det_Jacobian,cart_deriv] = B_matrix(posgp,n_unkn,n_nodes)
dshape = 0.25*[-1 -1;1 -1;1 1;-1 1]'.* flip(1+[-1 -1;1 -1;1 1;-1 1]'.*posgp,1);
Jacobian_mat = dshape*[0 0;1 0;1 1;0 1];
Det_Jacobian = det(Jacobian_mat);
cart_deriv = Jacobian_mat\dshape;
BE = zeros(3,n_unkn*n_nodes); BE([1 3],1:n_unkn:end) = cart_deriv; BE([3 2],2:n_unkn:end) = cart_deriv;
%%%% %%%% %%%% %%%% %%%% %%%% %%%% %%%% %%%%
function [DE] = D_matrix_stress(E,nu) %Planestress
DE = E/(1-nu^2)*[1 nu 0; nu 1 0;0 0 (1-nu)/2];
%%%% %%%% %%%% %%%% %%%% %%%% %%%% %%%% %%%%
function [coeff] = interp_property (m,n,beta,chi)
coeff = chi + (1-chi).*beta;
coeff = double(m==n).*coeff.^m + double(m~=n).*m*coeff.^n*(1-beta);
%%%% %%%% %%%% %%%% %%%% %%%% %%%% %%%% %%%%
function [K] = assmebly_stiff_mat (chi,KE,KE_cut,beta,m,iK,jK,n_unkn,nelx,nely)
sK = interp_property(m,m,beta,chi).*KE(:); sK(:,chi~=1&chi~=0) = interp_property(m,m,beta,chi(chi~=1&chi~=0)).*KE_cut(:);
K = sparse(iK,jK,sK,n_unkn*(1+nelx)*(1+nely),n_unkn*(1+nelx)*(1+nely)); K = (K+K')/2; clear sK;
%%%% %%%% %%%% %%%% %%%% %%%% %%%% %%%% %%%%
function [lambda,chi,psi,vol,Tol_constr] = find_volume (xi,connect,active_node,passive_node,t_ref,lambda,alpha0)
l1 = min(xi); c1 = t_ref; l2 = max(xi); c2 = t_ref-1; Tol_constr = 1; iter=1;
if lambda>l1 && lambda<l2
	[chi,psi,vol,l1,l2,c1,c2,Tol_constr] = compute_volume_lambda(xi,connect,active_node,passive_node,t_ref,lambda,l1,l2,c1,c2,alpha0);
end
while (abs(Tol_constr)>1e-4) && iter<1000
	lambda = 0.5*(l1+l2);
	[chi,psi,vol,l1,l2,c1,c2,Tol_constr] = compute_volume_lambda(xi,connect,active_node,passive_node,t_ref,lambda,l1,l2,c1,c2,alpha0);
	iter = iter + 1;
end
%%%% %%%% %%%% %%%% %%%% %%%% %%%% %%%% %%%%
function [chi,psi,vol,l1,l2,c1,c2,Tol_constr] = compute_volume_lambda(xi,connect,active_node,passive_node,t_ref,lambda,l1,l2,c1,c2,alpha0)
psi = xi - lambda; psi(passive_node) = -alpha0; psi(active_node) = alpha0;
[vol,chi] = compute_volume (psi,connect);
Tol_constr = -(vol-t_ref);
if Tol_constr > 0, l1 = lambda; c1 = Tol_constr; else; l2 = lambda; c2 = Tol_constr; end
%%%% %%%% %%%% %%%% %%%% %%%% %%%% %%%% %%%%
function [volume,chi] = compute_volume (psi,connect)
P = [-1 -1;1 -1;1 1;-1 1]; dvol = 1/4;
s = [-0.9324695142031521 -0.6612093864662645 -0.2386191860831969 0.2386191860831969 0.6612093864662645 0.9324695142031521]; [s,t] = meshgrid(s,s);
w = [ 0.1713244923791704  0.3607615730481386  0.4679139345726910 0.4679139345726910 0.3607615730481386 0.1713244923791704]; W=w'*w; W=W(:)';
psi_n = psi(connect); chi = sum((sign(psi_n)+1),2)'/8; id = chi~=1&chi~=0;
phi_x = psi_n(id,:)*((1+P(:,1)*s(:)').*(1+P(:,2)*t(:)')/4);
chi(1,id) = (W*(phi_x>0)'+ 0.5*W*(phi_x==0)')*dvol;
volume = 1 - sum(chi) / size(connect,1);
%%%% %%%% %%%% %%%% %%%% %%%% %%%% %%%% %%%%
function [fig_handle,obj_handle] = plot_isosurface(fig_handle,obj_handle,iter,psi,coord,connect,J,opt)
if opt.Plot_top_iso
	if iter==0; fig_handle = figure(2); set(fig_handle,'Name','Topology'); caxis([-1 1]); colormap(flip(gray(2)));
		axis equal tight; xlabel('x'); ylabel('y'); title(['J = ',num2str(J)]);
		obj_handle = patch('Vertices',coord,'Faces',connect,'FaceVertexCData',psi,'EdgeColor',opt.EdgeColor,'FaceColor','interp');
	else
		set(0, 'CurrentFigure', fig_handle); set(get(gca,'Title'),'String',['J = ',num2str(J)]);
		set(obj_handle,'FaceVertexCData',psi);
	end
end
%%%% %%%% %%%% %%%% %%%% %%%% %%%% %%%% %%%%
function [fig_handle,J_vec,vol_vec] = plot_volume_iter(fig_handle,iter,J_vec,J,vol_vec,vol,opt_plot,fig_num,fig_name,xlabel_name,linestyle)
J_vec = [J_vec,J]; vol_vec = [vol_vec,vol];
if opt_plot
	if iter==1; fig_handle = figure(fig_num); set(fig_handle,'Name',fig_name);
		subplot(2,1,1); plot(J_vec,linestyle);   ylabel('$\mathcal{J}_\chi$','Interp','Latex'); xlabel(xlabel_name); grid; grid minor;
		subplot(2,1,2); plot(vol_vec,linestyle); ylabel('|\Omega^-|');                          xlabel(xlabel_name); grid; grid minor;
	else; set(0, 'CurrentFigure', fig_handle);
		subplot(2,1,1); set(findobj(gca,'Type','line'),'Xdata',1:numel(J_vec),'YData',J_vec);
		subplot(2,1,2); set(findobj(gca,'Type','line'),'Xdata',1:numel(vol_vec),'YData',vol_vec);
	end
end
%%%% %%%% %%%% %%%% %%%% %%%% %%%% %%%% %%%%
function [vol] = set_reference_volume(iter,Vol0,Volf,nsteps,k)
if k==0; vol=Vol0+(Volf-Vol0)/nsteps*iter;
else; C1=(Vol0-Volf)/(1-exp(k)); C2=Vol0-C1; vol=C1*exp(k*iter/nsteps)+C2; end
\end{lstlisting}
		\end{strip}
	
\end{appendices}

\section*{Replication of results}
The Matlab codes provided in the paper, in Appendix \ref{app_matlab_code} and \href{https://github.com/DanielYago/UNVARTOP}{GitHub repository}, are the same ones used for obtaining the results here presented (Section \ref{sec_num_examples}). Therefore, they can be fully used as a replication tool, to reproduce those results, as well as to be used in additional numerical simulations. 

\bibliographystyle{abbrvnat}
\bibliography{document_StructMultiOpt}  

\end{document}